\documentclass[a4paper,11pt]{article}
\pdfoutput=1 
\usepackage{jcappub}
\usepackage{mathtools}
\usepackage{graphicx}
\usepackage{natbib}
\usepackage{multirow}
\usepackage{amsmath}
\usepackage[amssymb]{SIunits}
\usepackage{empheq}
\usepackage{booktabs}
\usepackage{slashed}
\usepackage{subfig}
\usepackage{bm}

\usepackage{color}

\newcommand{\rhodcdm}{\rho_{\rm dcdm}}
\newcommand{\rhodr}{\rho_{\rm dr}}

\newcommand{\deltadcdm}{\delta_{\rm dcdm}}
\newcommand{\thetadcdm}{\theta_{\rm dcdm}}
\newcommand{\deltadr}{\delta_{\rm dr}}
\newcommand{\thetadr}{\theta_{\rm dr}}
\newcommand{\mpsi}{\mathfrak{m}_{\psi}}
\newcommand{\mcont}{\mathfrak{m}_{\rm cont}}
\newcommand{\mshear}{\mathfrak{m}_{\rm shear}}

\newcommand{\fdcdm}{f_{\rm dcdm}}
\newcommand{\Gammadcdm}{\Gamma_{\rm dcdm}}

\usepackage[T1]{fontenc} 

\title{\boldmath Fractional Dark Matter decay: cosmological imprints and observational constraints}

\author[a,b]{Linfeng Xiao,}
\author[a,c]{Le Zhang,}
\author[a,b]{Rui An,}
\author[d]{Chang Feng,}
\author[e,f]{Bin Wang}

\affiliation[a]{Department of Astronomy, Shanghai Jiao Tong University, Shanghai, 200240, China}
\affiliation[b]{IFSA Collaborative Innovation Center, School of Physics and Astronomy, Shanghai Jiao Tong University, Shanghai 200240, China.}
\affiliation[c]{Shanghai Key Laboratory for Particle Physics and Cosmology, China}
\affiliation[d]{Department of Physics, University of Illinois at Urbana-Champaign, 1110 West Green Street, Urbana, Illinois, 61801, USA}
\affiliation[e]{School of Aeronautics and Astronautics, Shanghai Jiao Tong University, Shanghai 200240, China.}
\affiliation[f]{Center for Gravitation and Cosmology, YangZhou University, Yangzhou 225009, China.}

\emailAdd{hartley@sjtu.edu.cn}
\emailAdd{lezhang@sjtu.edu.cn}
\emailAdd{an\_rui@sjtu.edu.cn}
\emailAdd{changf@illinois.edu}
\emailAdd{wang\_b@sjtu.edu.cn}

\abstract{If a fraction $\fdcdm$ of the Dark Matter decays into invisible and massless particles (so-called ``dark radiation'') with the decay rate (or inverse lifetime) $\Gammadcdm$, such decay will leave distinctive imprints on cosmological observables.  With a full consideration of the Boltzmann hierarchy, we calculate the decay-induced impacts not only on the CMB but also on the redshift distortion and the kinetic Sunyaev-Zel'dovich effect, while providing detailed physical interpretations based on evaluating the evolution of gravitational potential. By using the current cosmological data with a combination of Planck 2015, Baryon Acoustic Oscillation and redshift distortion measurements which can improve the constraints, we update the $1\sigma$ bound on the fraction of decaying DM from $\fdcdm\lesssim5.26\%$ to $\fdcdm\lesssim2.73\%$ for the short-lived DM (assuming $\Gammadcdm/H_0\gtrsim10^4$). However, no constraints are improved from RSD data ($\fdcdm\lesssim0.94\%$) for the long-lived DM (i.e., $\Gammadcdm/H_0\lesssim10^4$). We also find the fractional DM decay can only slightly reduce the $H_0$ and $\sigma_8$ tensions, which is consistent with other previous works. Furthermore, our calculations show that the kSZ effect in future would provide a further constraining power on the decaying DM.}

\begin{document}
\maketitle
\flushbottom
\section{Introduction}
\label{sec:intro}

The $\Lambda$CDM model has been remarkably successful in accounting for a number of astronomical and cosmological observations of the universe such as the temperature and polarization anisotropies in the Cosmic Microwave Background (CMB) and the large scale structure. The existence of the Dark Matter (DM) in our universe is undoubtedly believed, whereas the particle physics nature of DM remains elusive after decades of research, which is mainly due to the fact that the cosmological observables are only sensitive to the purely gravitational effect of DM rather than its particle properties.

The recent Planck CMB measurements~\citep{Ade:2015xua,Aghanim:2018eyx} can achieve sub-percent accuracy in determining the cosmological parameters. However, the CMB results and other cosmological measurements have revealed several inconsistencies in the $\Lambda$CDM model, for instance, the $H_0$ tension~\cite{Riess:2011yx,Riess:2016jrr}, the $\sigma_8$ tension~\cite{Ade:2015xua,Hamann:2013iba,Battye:2013xqa,Petri:2015ura} and the missing satellite problem~\cite{Klypin:1999uc,Simon:2007dq}, etc. On the theory side, the cosmological constant problem~\cite{Weinberg:1988cp} and the coincidence problem~\cite{Chimento:2003iea} remain the major challenges in the modern cosmology. Many attempts have been made so far to remedy the tensions on $H_0$ and $\sigma_8$ by taking account of some new physics beyond the $\Lambda$CDM model, for instance, a holographic dark energy plus sterile neutrino model~\cite{Guo:2018ans}, non-zero coupling in the dark sector components~\cite{Kumar:2019wfs}, a minimally coupled and slowly-or-moderately rolling quintessence field~\cite{Miao:2018zpw}, multiple Dark Energy (DE) models~\cite{Mortsell:2018mfj} and the early dark energy~\cite{Poulin:2018cxd}, or even some models in the form of phantom DE or extra radiation~\cite{Vagnozzi:2019ezj}. Alternatively, one can take into account modifications of the standard cold DM scenario, including a cannibal dark matter~\cite{Buen-Abad:2018mas}, partially acoustic dark matter models~\cite{Raveri:2017jto}, dissipative dark matter models~\cite{daSilva:2018dzn}, hot axions~\cite{DEramo:2018vss}, charged DM with chiral photons~\cite{Ko:2017uyb} and decaying DM scenarios~\cite{Doroshkevich:1989bf,Kaplinghat:1999xy,Ichiki:2004vi,Lattanzi:2007ux,Gong:2008gi,DeLopeAmigo:2009dc,Peter:2010au,Peter:2010sz,Huo:2011nz,Aoyama:2011ba,Wang:2013rha,Aoyama:2014tga,Audren:2014bca,Chudaykin:2016yfk,Poulin:2016nat,Chudaykin:2017ptd}, etc., which provide other possible solutions to resolve the tensions. Moreover, DM self-interactions~\cite{Tulin:2017ara} and DM-DE interacting models~\cite{Amendola:1999er,Amendola:2003eq,Pavon:2005yx,Boehmer:2008av,Olivares:2006jr,Bertolami:2007zm,Micheletti:2009pk,Honorez:2010rr,Costa:2014pba,vandeBruck:2015ida,Wang:2016lxa,Costa:2016tpb,Yang:2018euj} as well as the modified gravity~\cite{DeFelice:2010aj,Harko:2014gwa} are also possible solutions to fix such the tensions.

On the other hand, a variety of hypothesized extensions of the Standard Model of particle physics generally predict new dynamics between the electroweak and the Planck scales together with a number of new particles readily with the required properties to be dark matter. For instance, the extremely light axions~\cite{Preskill:1982cy,Peccei:1977hh}, the Weakly Interacting Massive Particles (WIMPs)~\cite{Goldberg:1983nd,Jungman:1995df,Lee:1977ua} and the Kaluza-Klein DM~\cite{Servant:2002aq,Cheng:2002ej} are most popular DM scenarios, which not only offer solutions to several shortcomings of the Standard Model, but also provide possibilities to solve the observed cosmological tensions. If DM is composed of these particles, it could be unstable and can decay on cosmological time-scales, yielding visible effects in cosmological observations. The influences of unstable DM producing electromagnetically-interacting particles on the CMB and the large scale structure are first investigated by~\cite{Adams:1998nr,Chen:2003gz,Zhang:2007zzh,Zhang:2006fr}.

In this work, we consider a fractional cold DM decay model~\cite{Poulin:2016nat} that allows to mitigate the $H_0$ and $\sigma_8$ tensions~\cite{Pandey:2019plg,Vattis:2019efj,Enqvist:2015ara}. The decay products are invisible and massless particles, i.e, ``Dark Radiation'' (DR) and can not be identified/constrained by cosmic-ray measurements. In this study, we therefore investigate its cosmological imprints on CMB temperature anisotropies and the large-scale structure growth as well as the baryon velocity field via the evolution of the gravitational potential in terms of this DM decay model. We perform a joint analysis by combining the CMB datasets from Planck 2015~\cite{Ade:2015xua}, Baryon Acoustic Oscillation (BAO) and the redshift distortion (RSD) data at very low redshifts. Furthermore, we propose the use of the kinetic Sunyaev-Zel'dovich (kSZ) effect to probe/constrain the DM decay model from further surveys.

The paper is organized as follows. In Sec.~\ref{sec:Equations} we describe our fractional DM decay model and present the basic equations for the evolution of the background dynamics and the linear perturbations. In Sec.~\ref{sec:effects}, we examine the effects of the decaying DM model on various cosmological observables, and we present our results in Sec.~\ref{sec:results}. Sec.~\ref{sec:conclu} is devoted to conclusions.

\section{A fractional DM decay model}
\label{sec:Equations}
Here we briefly describe the main equations describing the DM-decay induced gravitational impacts by following \cite{Poulin:2016nat}, assuming that the fractional DM decays into invisible and massless particles, dubbed as ``Dark Radiation'' (DR). In this DM-decaying model, the ratio of the decaying cold DM (DCDM) fraction to the total one, $\fdcdm$, and the decay rate,  $\Gammadcdm$, are two free parameters. With these notations, the abundance of the standard cold DM (SDM) is then $1-\fdcdm$. With the standard procedure, we split up the evolution equations into the homogeneous background part and the small perturbation part as follows.

\subsection{Background equations}

With a considerion of DM decay, energy conservation for the DM and dark radiation in the unperturbed background metric yields
\begin{eqnarray}
\rhodcdm '=-3\frac{a'}{a}\rhodcdm-a\Gammadcdm\rhodcdm~,\label{eq:rhodcdm}\\
\rhodr '= -4\frac{a'}{a}\rhodr + a\Gammadcdm\rhodcdm~.\label{eq:rhodr}
\end{eqnarray}
The prime (') above, and thereafter denotes the derivative with respect to conformal time, and $a$ is the scalar factor of the universe. Here $\rhodcdm$ denotes the density of the fractional decaying DM (i.e., $\rhodcdm\equiv \fdcdm \cdot \rho_c$, where $\rho_{c}$ denotes the total DM density), and $\rhodr$ is the density of dark radiation. Here we assume a positive $\Gammadcdm$ that implies the decays from DM to DR, but not vice verse.  We shall come back to the standard $\Lambda$CDM model if either $\Gammadcdm$ or $\rhodr$ is set to be zero.

\subsection{Perturbation evolutions}

After having computed the cosmological perturbations at the linear order~\citep{Audren:2014bca} and following the notation of Ref.~\citep{Poulin:2016nat}, one can derive the expressions of the DCDM scalar perturbations in both synchronous and Newtonian gauges, which read

\begin{eqnarray}\label{eq:ContinuityDecay}
\deltadcdm'&=&-\thetadcdm-\mathfrak{m}_{\rm{cont}}-a\Gamma\mathfrak{m}_{\psi}~,\\
\thetadcdm'&=&-\mathcal{H}\thetadcdm+k^2\mathfrak{m}_{\psi}~,
\end{eqnarray}
where $\delta_{\rm  dcdm}$ is the DCDM overdensity and $\thetadcdm$ is the velocity divergence, $k$ the wavenumber in Fourier space and $\mathcal{H} = \frac{a'}{a}$. Note that, as proposed in Ref.~\cite{Audren:2014bca}, we have introduced two variables $\mathfrak{m}_{\rm{cont}}$ and $\mathfrak{m}_{\psi}$. Since the numerical code CAMB~\cite{Lewis:1999bs} is written in the synchronous gauge and the Newtonian gauge is convenient for later discussions on physical aspects, we provide the expression of the sources terms in both gauges, reported in Tab.~\ref{table:metricSourceTerms}.

Furthermore, by following Ref.~\cite{Ma:1995ey,Poulin:2016nat}, the full Boltzmann hierarchy to describe the perturbations induced by DR reads

\begin{eqnarray}\label{eq:HierarchyWithDecayBothGauges}
F_{\rm dr,0}' & = & -kF_{{\rm dr},1}-\frac{4}{3}r_{\rm dr}\mcont+r_{\rm dr}'(\delta_{\rm dcdm}+\mpsi)~,\\
F_{\rm dr,1}' & = & \frac{k}{3}F_{{\rm dr},0}-\frac{2k}{3}F_{{\rm dr},2}+\frac{4k}{3}r_{\rm dr}\mpsi+\frac{r_{\rm dr}'}{k}\theta_{\rm dcdm}~,\\
F_{\rm dr,2}' & = & \frac{2k}{5}F_{{\rm dr},1}-\frac{3k}{5}F_{{\rm dr},3}+\frac{8}{15}r_{\rm dr}\mshear~,\\
F_{\rm dr,\ell}' & = & \frac{k}{2\ell+1}\big(\ell F_{{\rm dr},\ell-1}-(\ell+1)F_{{\rm dr},\ell+1}\big)\qquad \ell>2\,,
\end{eqnarray}
which is the gauge-independent hierarchy and has to be truncated at some large multipole $\ell$ for calculation, and the anisotropic stress developing in the fluid is governed by the lowest three multipoles $F_{{\rm dr},\ell}$, which in terms of the standard variables $\delta$, $\theta$, $\sigma$ in DR fluid read: 
\begin{eqnarray}\label{eq:FtoStandardVariables}
F_{{\rm dr},0} = r_{{\rm dr}}\delta_{{\rm dr}}~,\qquad\qquad F_{{\rm dr},1} = \frac{4r_{{\rm dr}}}{3k}\theta_{{\rm dr}}~,\qquad\qquad  F_{{\rm dr},2}=2\sigma r_{{\rm dr}}\,.
\end{eqnarray}
Here, $r_{\rm dr}$ is defined as
\begin{equation}
r_{\rm dr}\equiv\frac{\rho_{\rm dr}a^4}{\rho_{cr,0}}\,,
\end{equation}
with $\rho_{cr,0}$ to be the critical density at present. The conformal time derivative of $r_{\rm dr}$ is expressed by 
\begin{eqnarray}
r_{\rm dr}' & = & a\Gammadcdm \frac{\rhodcdm}{\rho_{\rm dr}}r_{\rm dr}\,.
\end{eqnarray}
 By using Eq.~\ref{eq:FtoStandardVariables}, one would arrive at the final set of equations:
\begin{eqnarray}
\deltadr'&=&-\frac{4}{3}(\thetadr+\mcont)+a\Gammadcdm\frac{\rhodcdm}{\rhodr}(\deltadcdm-\deltadr+\mpsi)~,\label{eq:delta_DR}\\
\thetadr'&=&\frac{k^2}{4}\deltadr-k^2\sigma_{\rm dr}+k^2\mpsi-a\Gammadcdm\frac{3\rhodcdm}{4\rhodr}\bigg(\frac{4}{3}\thetadr-\thetadcdm\bigg)~.\label{eq:theta_DR}
\end{eqnarray}

Note that, Eqs.~\ref{eq:delta_DR} and \ref{eq:theta_DR} are rather similar, but not exactly equivalent to those for massless neutrino due to the decay terms as the function of $\Gammadcdm$. 

\begin{table}[!h]
\centering
\begin{tabular}{l|cc}
	& Synchronous \quad& Newtonian\\
	\hline
   $\mathfrak{m}_{\rm{cont}}$  & $h'/2$ & $-3\phi'$ \\
   $\mathfrak{m}_{\psi}$ & 0& $\psi$ \\
   $\mshear$  & $(h'+6\eta')/2$ & 0 \\
 \end{tabular}
  \caption{\label{table:metricSourceTerms} Three metric source terms for scalar perturbations and their expressions in synchronous and Newtonian gauges.}
 \end{table}

\section{Cosmological Imprints from fractional DM decay}\label{sec:effects}
In this section, we investigate the purely DM-decaying induced gravitational effects on the CMB temperature anisotropies, redshift distortion (RSD) and kinetic Sunyaev-Zel'dovich effect (kSZ), as a function of the decaying DM fraction, $\fdcdm$, and its decay rate, $\Gammadcdm$. The imprints on the CMB and the matter power spectrum has been already discussed in Ref.~\citep{Poulin:2016nat}, and we in this study further calculate the DM-decaying induced effects on the RSD and kSZ, which would be measured accurately in future observations. Before doing so, it is worthwhile clarifying the initial condition setting for our DCDM model and demonstrating DCDM background evolution.

\subsection{Initial Condition Setting and DCDM Background Evolution}\label{sec:ICS_BG}

In order to illustrate the effect of the DM decay rate on the CMB temperature angular power spectrum and to qualitatively discuss the corresponding physical origins clearly, we have to fix some cosmological parameters in calculations. As known, if we were altering the decay rate $\Gammadcdm$ while keeping the DM density remains same as that at the present day, the solver CAMB would automatically adjust the initial conditions in the early universe. However, such adjustments would lead to the effects of $\Gammadcdm$ on the perturbations mixed with that of changing the early cosmological evolution.

Therefore, a better choice (as in Ref.~\cite{Audren:2014bca,Poulin:2016nat}) is to fix the following cosmological parameters at the early time of $a=10^{-8}$ same as those in standard $\Lambda$CDM universe, including the baryon density $\rho_b$, the total DM density $\rho_c$, the amplitude of primordial perturbation $A_s$, the index of the primordial perturbation spectrum $n_s$, the redshift of reionization $z_{\rm reio}$ and the angular size of the sound horizon $\theta_{\rm MC}$. We also assume that the decaying DM of the density $\fdcdm\cdot\rho_c$ starts to decay at $a=10^{-8}$. With these parameters fixed and varying $\Gammadcdm$ and $\fdcdm$, as the consequence, the present-day value of DM density $\rho_c$, the optical depth $\tau_{\rm reio}$ and the Hubble constant $H_0$ will change accordingly. These fixed parameters are chosen from Planck 2015 best-fitted TT, TE, EE+Low-P parameters~\cite{Ade:2015xua}: \{$\Omega_bh^{2}=0.02225$, $\Omega_ch^{2}=0.1198$, $100\theta_{\rm MC}=1.04077$, $ln(10^{10}A_s)=3.094$, $n_s=0.9645$, $z_{\rm reio}=9.9$ and $h=0.6727$\}. To avoid confusions, hereinafter we abbreviate $\omega_b^{ini}=\Omega_bh^{2}$, $\omega_c^{ini}=\Omega_ch^{2}$ and $H_{0}^{ini}=100h$ km/s/Mpc as the input/initial parameters. {\it Note that, the angular size of the sound horizon $\theta_{\rm MC}$ is fixed in this section such that the peak positions of CMB TT spectrum are unaltered, which is convenient for our discussions and can cancel the trivial effects in the CMB~\cite{Poulin:2016nat}}. When performing the Monte-Carlo analysis in Sec.~\ref{sec:results}, we of course leave $\theta_{\rm MC}$ as a parameter. For illustration purpose, we fix the fraction of decaying DM to be $20\%$ and the fiducial decay rate $\Gammadcdm$ \footnote{To be consistent with the units used in CAMB, the decay rate in this study is expressed in units of ${\rm Mpc^{-1}}$, related to other works making use of $\rm{Gyr^{-1}}$ by the conversion factor of ${\rm Mpc^{-1}} = 307.22 \rm{Gyr^{-1}}$} in the range of $3\times10^{-4}$ -- $3\times10^3$ Mpc$^{-1}$. The evolution equations are numerically solved by a modified CAMB~\cite{Lewis:1999bs} code.

\begin{figure}
\centering
\includegraphics[scale=0.5]{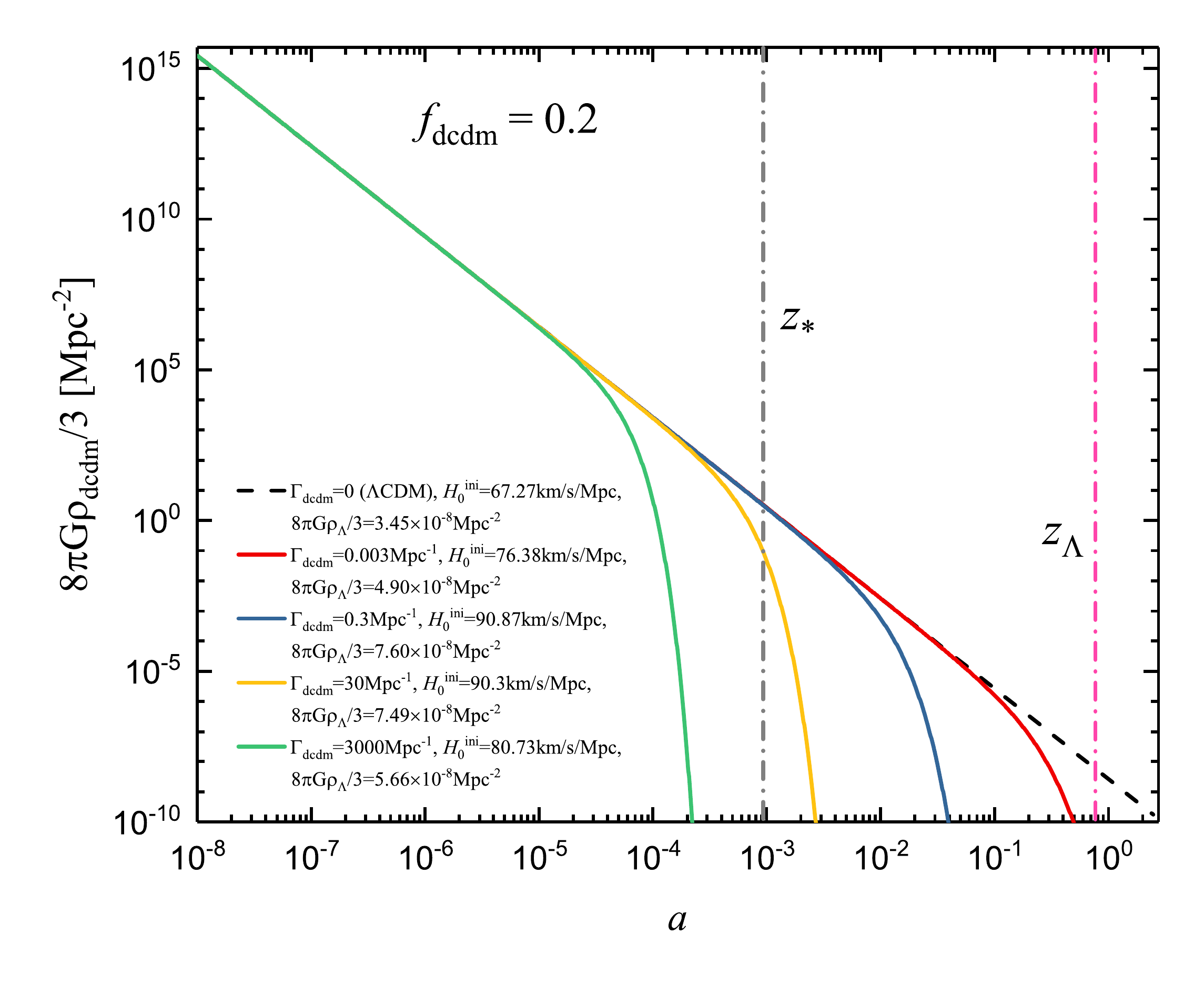}
\caption{\label{fig:DCDM_BG} DCDM background evolution for several decay rates with  $\fdcdm = 0.2$ fixed. The vertical lines indicate the typical times of the photon decoupling ($z_*\simeq 1089$) and when the universe becomes $\Lambda$-dominated ($z_\Lambda\simeq0.4$) in the standard $\Lambda$CDM model. Compared with the standard $\Lambda$CDM model (the black dashed line) without any decaying DM (i.e., $\Gammadcdm=0$), the density of short-lived DCDM ($\Gammadcdm \gg 0.3$ Mpc$^{-1}$) decreases significantly during the recombination epoch and the decrease of the density for the long-lived ($\Gammadcdm \lesssim 0.3$ Mpc$^{-1}$) one occurs at the late-time universe, as expected.}
\end{figure}

In Fig.~\ref{fig:DCDM_BG}, we illustrate the evolution of the background DCDM density evolution, starting with our initial conditions. We observe that, for the long-lived DCDM where $\Gammadcdm \lesssim 0.3$ Mpc$^{-1}$ (i.e., longer than recombination time), the evolution starts to deviate from the standard $\Lambda$CDM one when $z\gtrsim z_*$, which would lead to different predictions for $H_0$, $\sigma_8$ and be able to mitigate their tensions from recent cosmological data. For very large $\Gammadcdm$ ($\Gammadcdm \gg 0.3$ Mpc$^{-1}$), corresponding to very short-lived DCDM, the DCDM density decreases significantly well before recombination ($z_*$), leaving strong imprints on the CMB.

In all cases, $\rho_ca^3$ eventually become smaller than its initial value due to the DCDM decay so that $\rho_\Lambda$ is automatically adjusted to higher values in order to keep the angular size $\theta_{\rm MC}= d_s/d_A$ fixed, in which $d_s$ is the sound horizon before photon decoupling and $d_A$ is the angular diameter distance to the last scattering surface. As a result, $H_{0}^{ini}$ (or equivalently the $\Lambda$ density $\rho_{\Lambda}$) has to be changed and depend on $\Gammadcdm$ non-monotonically when varying the decay rate. See more details and discussions in Appendix~\ref{sec:app_A}.

\subsection{Impacts on the CMB TT spectrum from DM decay}
\label{sec:CMB}

\begin{figure}[htbp!]
\centering
\subfloat[]{
\includegraphics[scale=0.32]{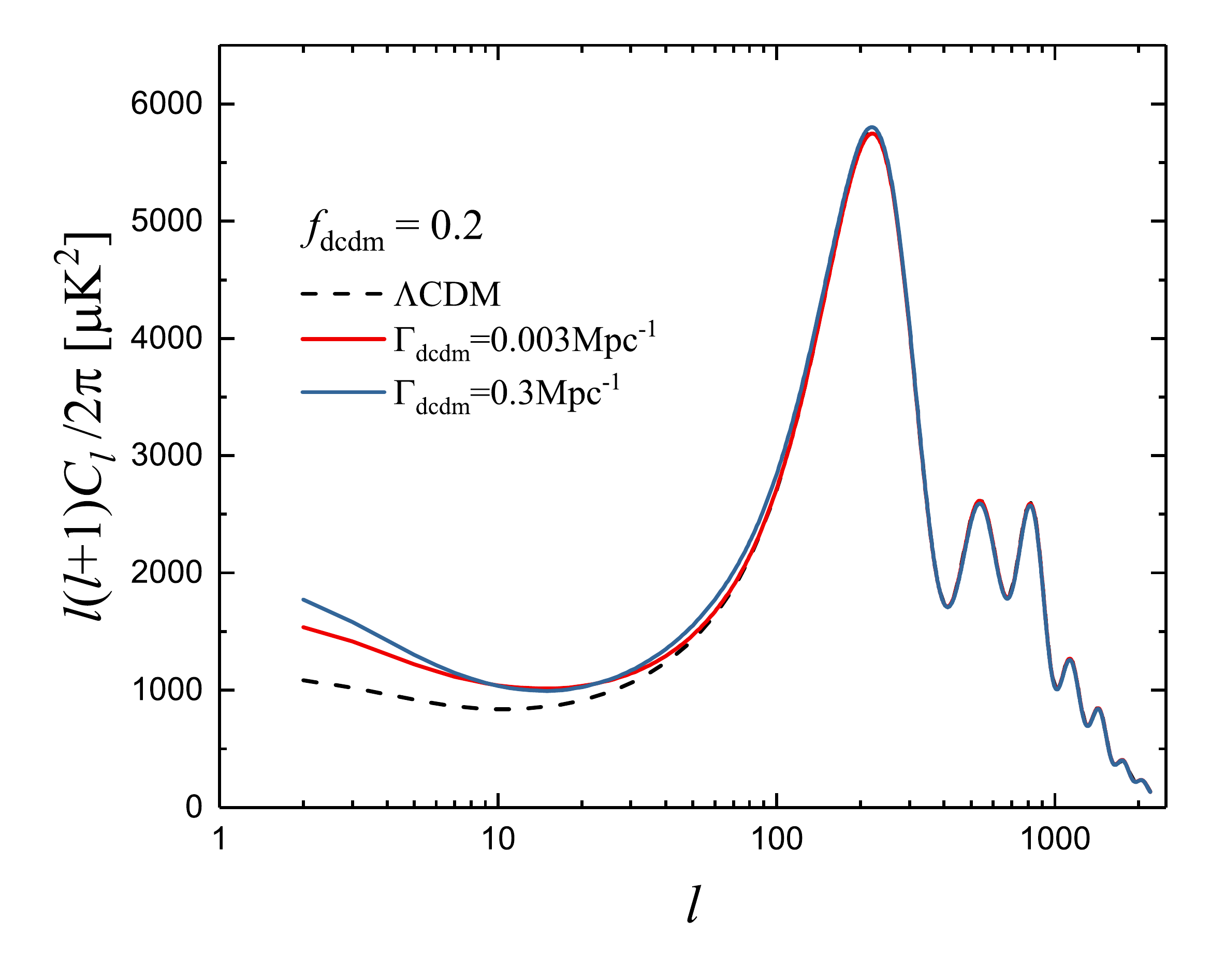}\label{fig:CMB1}}
\subfloat[]{
\includegraphics[scale=0.33]{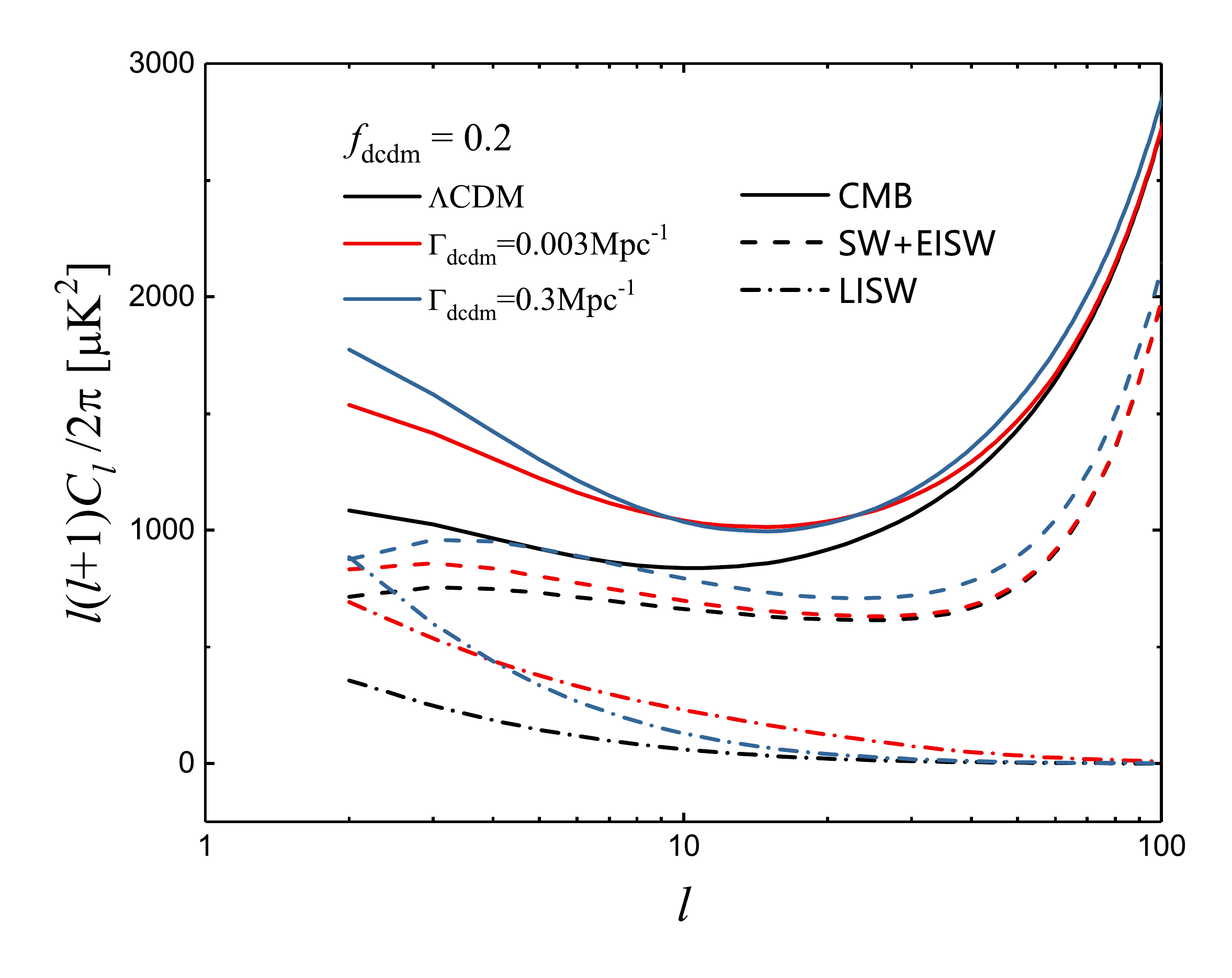}\label{fig:SW1}}
\caption{\label{fig:CMB_Gamma1}(a) TT angular power spectrum for the $\Lambda$CDM and the long-lived DCDM model. The long-lived DCDM mainly leaves imprints on the CMB spectrum at $\ell\lesssim 100$, boosting the amplitude at very large scale $\ell(\lesssim 10)$. (b) The imprints of the DM decay on the power spectra from the SW, early-ISW and late-ISW at $\ell\leqslant 100$, respectively.} 
\end{figure}

\begin{figure}[htbp!]
\centering
\subfloat[]{
\includegraphics[scale=0.33]{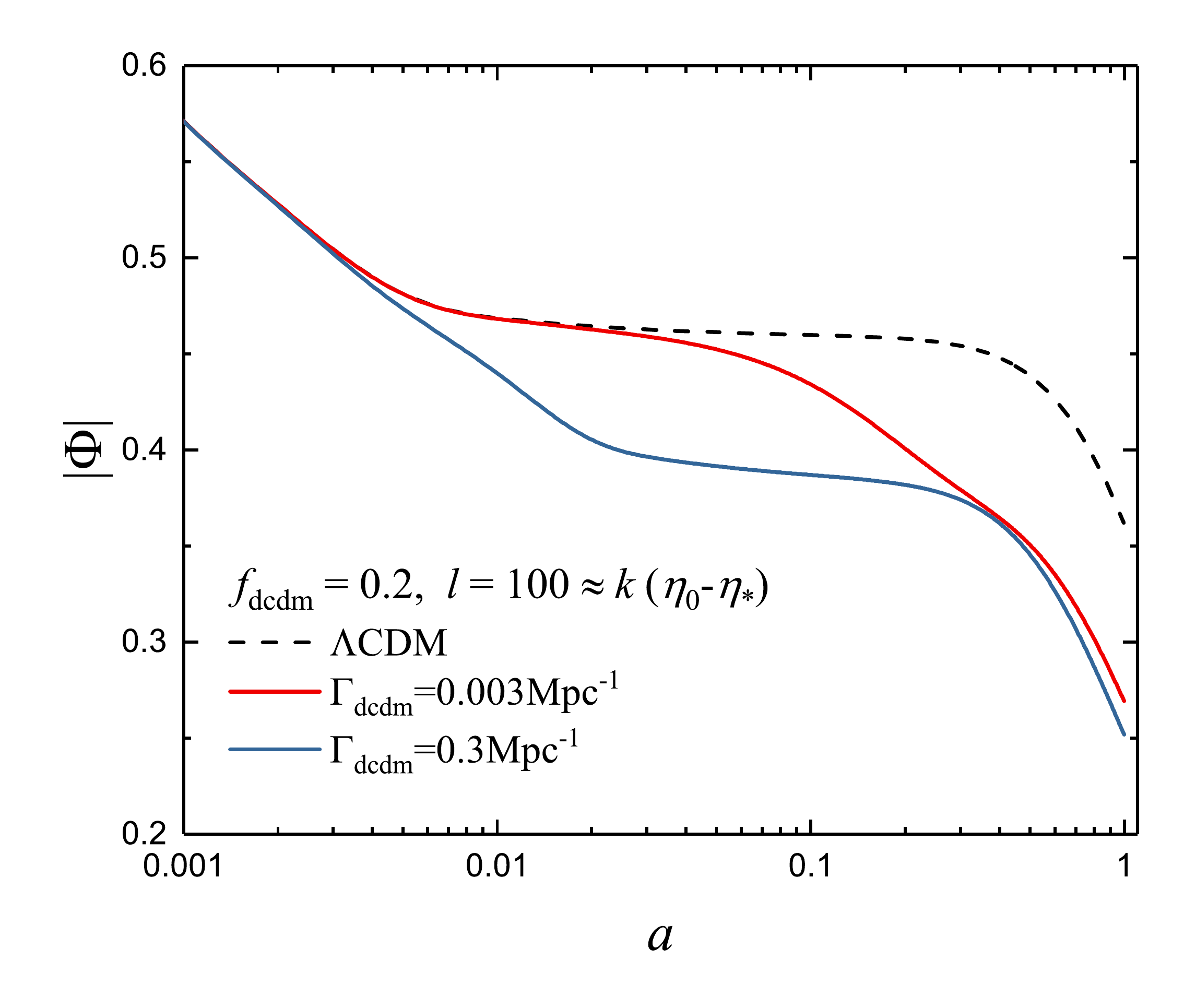}\label{fig:Phi1_l100}}
\subfloat[]{
\includegraphics[scale=0.33]{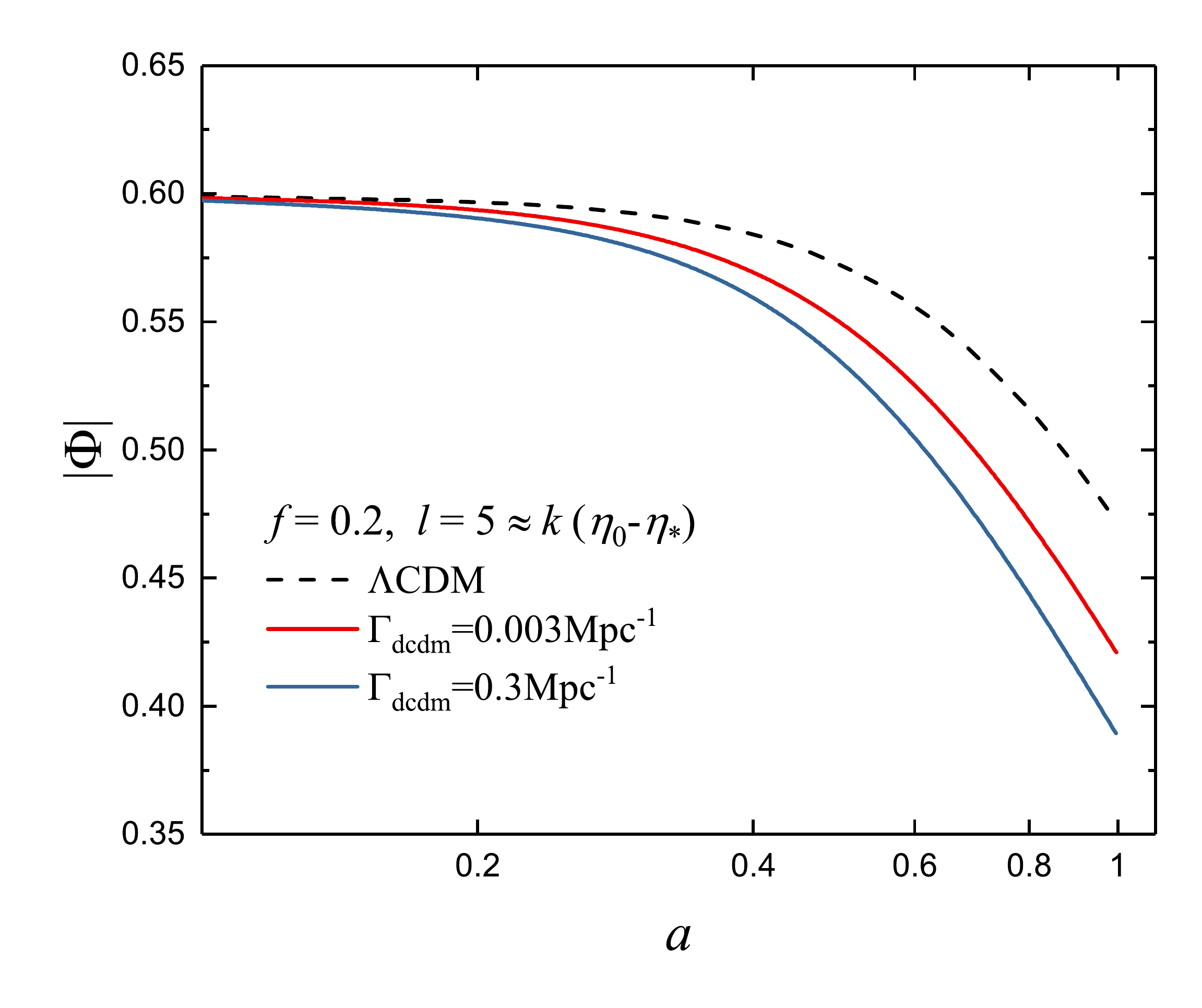}\label{fig:Ph11_l5}}
\caption{\label{fig:Phi1}The evolution of Weyl potential for the $\Lambda$CDM and DCDM models with two specific decay rate: $\Gammadcdm=0.003$ Mpc$^{-1}$ and $\Gammadcdm=0.3$ Mpc$^{-1}$. (a) The evolution of $|\Phi|$ at a fixed wavenumber of $k\approx 100/(\eta_0-\eta_*)$ after photon decoupling. (b) The evolution of $|\Phi|$ evolution at $k\approx 5/(\eta_0-\eta_*)$ during late universe. Larger potential suppression would lead to a stronger enhancement in SW and ISW effects, as shown in Fig.~\ref{fig:CMB_Gamma1}.} 
\end{figure}

In this subsection we investigate the impacts of the decay on the CMB TT angular power spectrum. In order to fully understand the changes in CMB power spectrum, we will carefully examine the decay-induced impacts on the Sachs-Wolfe effect (SW), the early Integrated Sachs-Wolfe effect (early-ISW) and the late Integrated Sachs-Wolfe effect (late-ISW), as well as on the gravitational potential.


Let us first focus on the long-lived DCDM models where $\Gammadcdm\lesssim 0.3$ Mpc$^{-1}$. From Fig.~\ref{fig:CMB1}, the amplitudes of the power spectrum at $\ell\gtrsim 100$ are almost identical for the different $\Gammadcdm$, but the decay apparently increases the amplitude at low-$\ell$ regime, $\ell\lesssim 10$.

These features are consistent with what we expect in the long-lived case from Fig.~\ref{fig:DCDM_BG}, since the DM lifetime longer than the recombination time would be almost no effect in the early universe but change the late universe considerably. The CMB perturbation modes at small scales with $\ell\simeq100$ have already well entered the horizon after recombination, so that the decay is not able to affect those modes. As we fix $\theta_{\rm MC}$ in this section and decay will lead to a decrease in $\rho_c$, the density $\rho_\Lambda$ has to be adjusted to a higher value when decay occurs (see Appendix~\ref{sec:app_A} in detail). This process will always compensate for decreasing $\rho_c$ by increasing $\rho_{\Lambda}$. As a result, a higher value of $\rho_\Lambda$ can lead to an enhancement in the ISW effect and consequently in low-$\ell$ CMB amplitudes.


The enhancement in low-$\ell$ amplitude in fact comes from the decay-induced changes in the SW, early-ISW and late-ISW effects, which is shown in Fig.~\ref{fig:SW1}. Let us focus on two cases: $\Gammadcdm =0.3$ Mpc$^{-1}$ and $0.003$ Mpc$^{-1}$, respectively. We find that, the decay will increase the both SW and early-ISW effects at $\ell<100$ and the larger decay rate increase the total amplitude of the SW and early-ISW more than from the smaller one, whereas the increase in the late-ISW from the smaller one could be greater than that from the larger one, except for at very large scales of $\ell\lesssim4$ where the contribution to late-ISW is mainly from incerasing $H_0^{ini}$ (i.e., increasing $\rho_{\Lambda}$ in order to keep $\theta_{\rm MC}$ fixed).

In order to physically understand these observed behaviors, we will investigate the evolution of the gravitational potential, namely the Weyl potential $\Phi = (\phi+\psi)/2$ defined in CAMB, where the scalar metric perturbations $(\phi,\psi)$ evaluated in the Newtonian gauge are shown in Table \ref{table:metricSourceTerms}. Intuitively due to the fact that the decay will decrease $\Omega_c$, we expect the growth of $\Phi$ would be suppressed. As known, the SW and ISW effects are determined by the evolution of time-dependent gravitational potential. In Fig.~\ref{fig:Phi1}, we illustrate the potential $|\Phi|$ for given wavenumbers $k$ as a function of scalar factor $a$. The wavenumber $k$ is related to the multipole $\ell$ through the relation of $k\approx \ell/(\eta_0-\eta_*)$, where $\eta_*$ and $\eta_0$ denote the conformal time at the recombination and present day, respectively. As seen, compared with the standard $\Lambda$CDM case, the smaller decay rate $\Gammadcdm=0.003$ Mpc$^{-1}$ mainly affect the potential $|\Phi|$ at $a\gtrsim0.02$ and the larger rate $\Gammadcdm=0.3$ Mpc$^{-1}$ affects $\Phi$ earlier. Such influences increase the changes of $|\Phi|$ after recombination, leading to a strong boost in the SW and ISW effects. Moreover, in Fig.~\ref{fig:SW1}, the larger decay rate $\Gammadcdm=0.3$ Mpc$^{-1}$ can offer a strong SW and early-ISW effects at $\ell\sim100$. For the late-ISW effect at $\ell=5$ in Fig.~\ref{fig:SW1}, the potential decay at $a\gtrsim 0.1$ in DCDM model is stronger than that in the standard $\Lambda$CDM model (seen from Fig.~\ref{fig:Ph11_l5}) and the potential $|\Phi|$ become lower for a larger $\Gammadcdm$, as expected.

From the above discussion about the case of the small decay rate, we are convinced that, the long-lived DCDM has almost no effect in the TT power spectrum at $\ell\gtrsim 100$ whereas it increases the amplitude of the spectrum notably at very large scales of $\ell\lesssim 10$. Note that, again, the angular size $\theta_{\rm MC}$ under the different decay rates is fixed to the value derived from the standard $\Lambda$CDM model in the calculations, and therefore the density $\rho_{\Lambda}$ has to be adjusted accordingly.

\begin{figure}[htbp!]
\centering
\subfloat[]{
\includegraphics[scale=0.32]{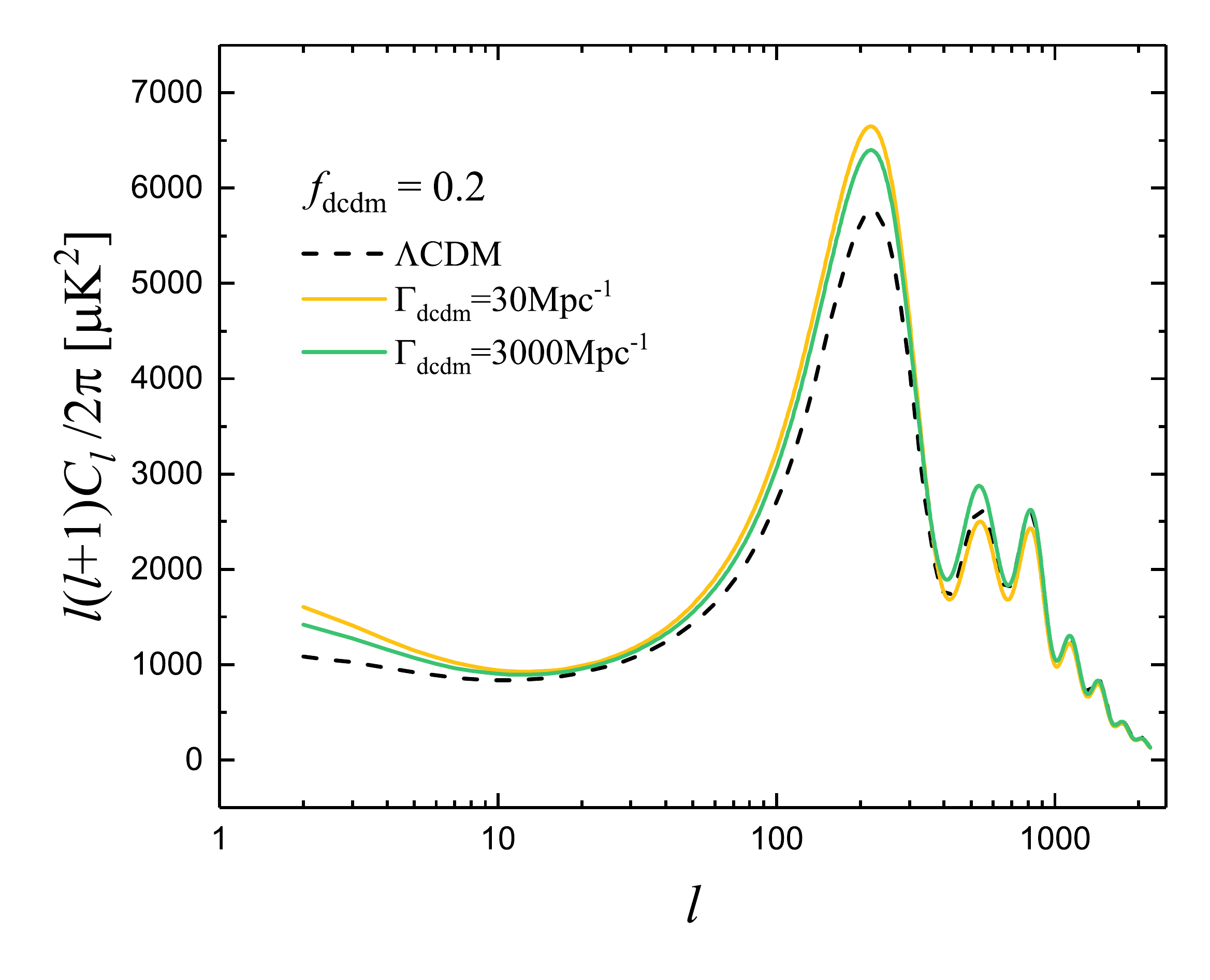}\label{fig:CMB2}}
\subfloat[]{
\includegraphics[scale=0.33]{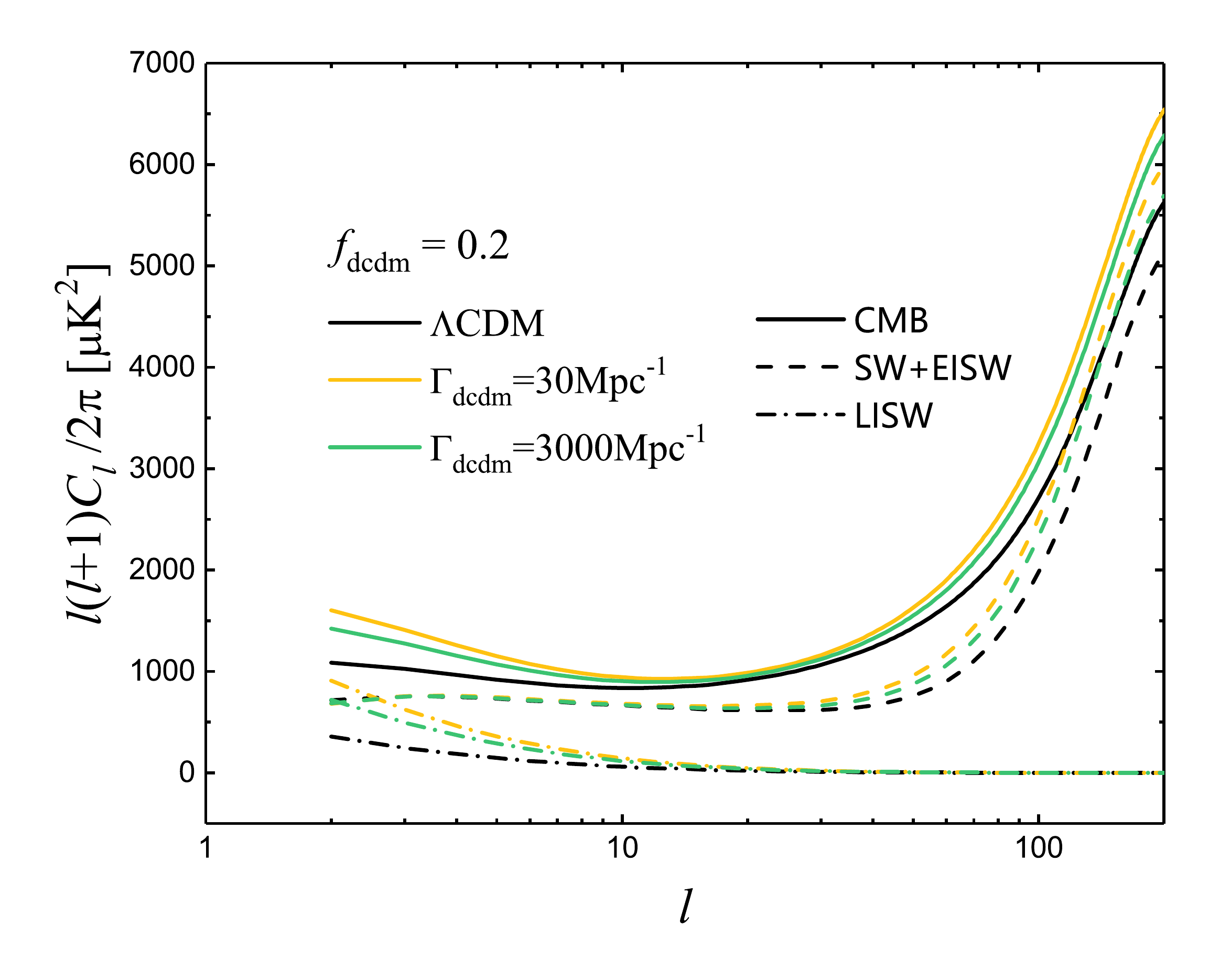}\label{fig:SW2}}
\caption{\label{fig:CMB_Gamma2} Same as in Fig.~\ref{fig:CMB_Gamma1}, but for several large decays rates, corresponding to the short-lived scenario. } 
\end{figure} 

\begin{figure}[htbp!]
\centering
\subfloat[]{
\includegraphics[scale=0.33]{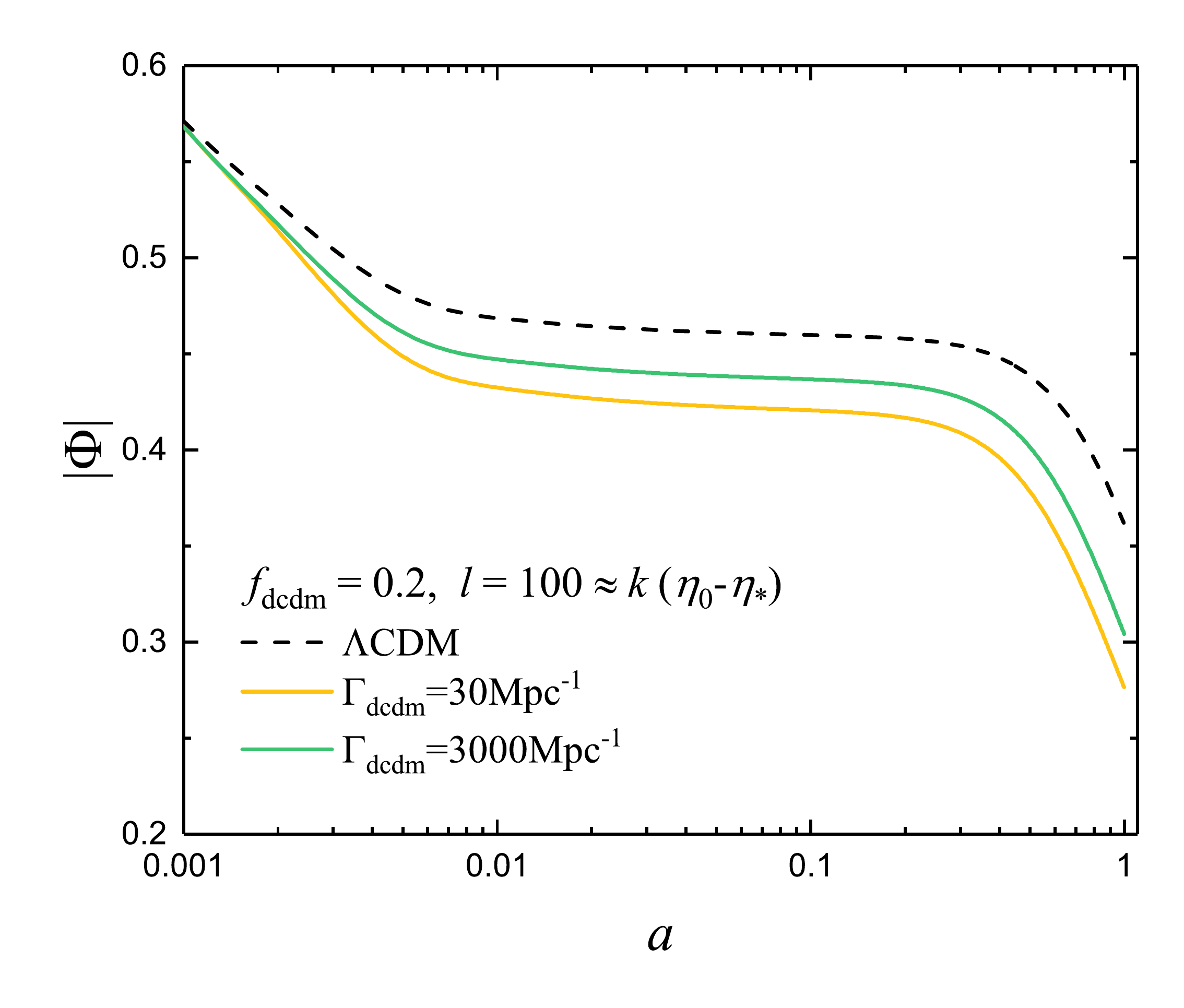}\label{fig:Phi2_l100}}
\subfloat[]{
\includegraphics[scale=0.33]{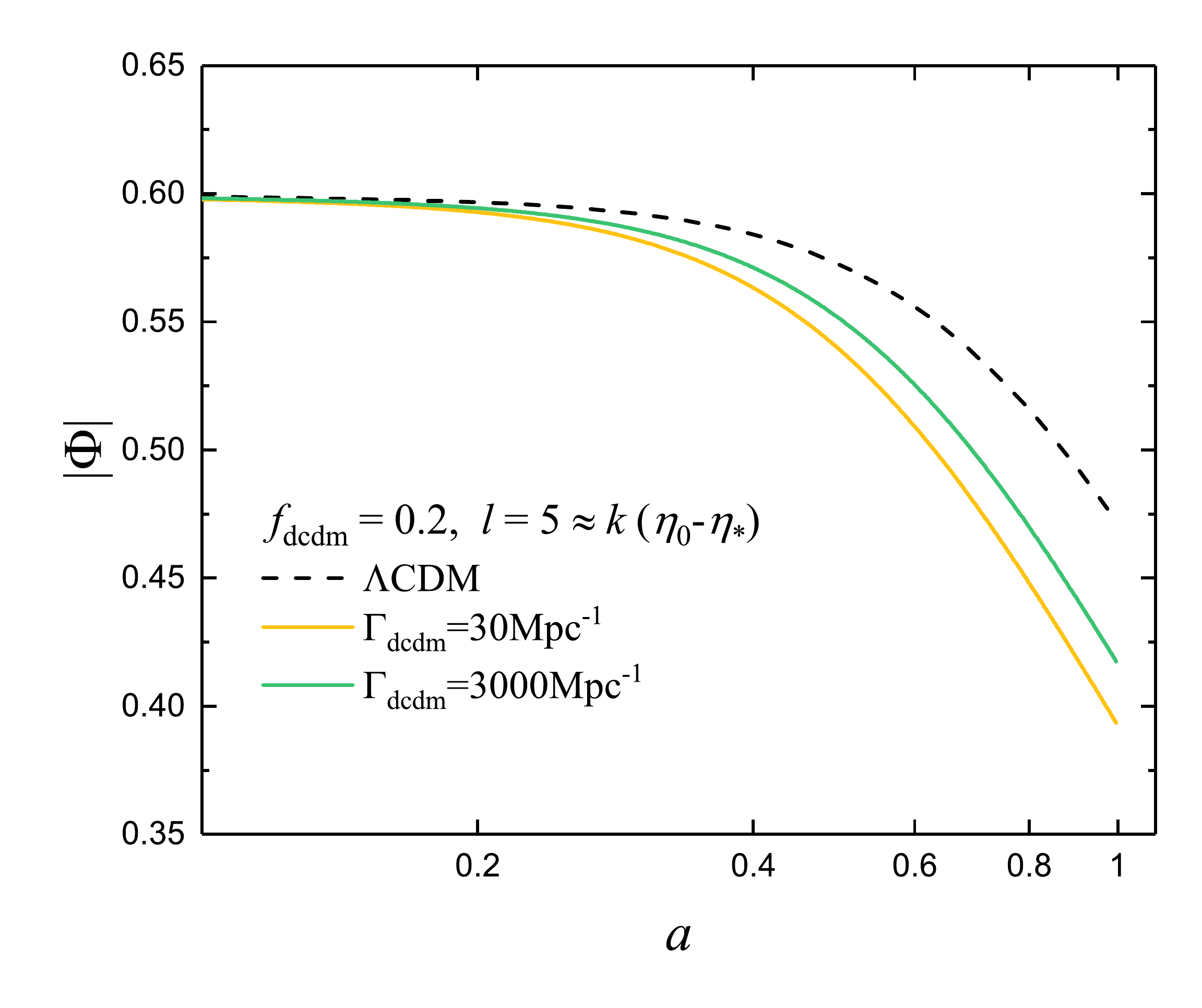}\label{fig:Ph12_l5}}
\caption{\label{fig:Phi2}Same as in Fig.~\ref{fig:Phi1}, but for the decay rates of $\Gammadcdm=30$ Mpc$^{-1}$ and $3000$ Mpc$^{-1}$. Compared with the standard $\Lambda$CDM model, the decay-induced deviations occur at the early universe before the recombination. A detailed interpretation of the dependence of $|\Phi|$ on $\Gammadcdm$ is presented in Appendix.~\ref{sec:app_B}.} 
\end{figure}

Next, let us move to the short-lived case which has a large decay rate with $\Gammadcdm>0.3$ Mpc$^{-1}$. As expected, the short-lived DCDM would play an important role at early universe before recombination, and leads to considerable changes in the CMB power spectrum at a broad range of multipoles, $\ell\lesssim 800$, shown in Fig.~\ref{fig:CMB_Gamma2}. To understand these features, following the same line of reasoning as in small $\Gammadcdm$ case, Fig.~\ref{fig:SW2} and ~\ref{fig:Phi2} show the SW and the early/late-ISW effects, and the evolution of  $|\Phi|$, respectively. As seen, due to the decay of DM which leads a suppression in $|\Phi|$, both the SW and ISW effects are enhanced visibly at the scales of $\ell\leq 200$. However, contrary to an intuitive expectation that the changes in $\Phi$ from smaller decay rates are weaker than from larger ones, the smaller decay rate $\Gammadcdm$ in fact will lead to a stronger influence in the CMB power spectrum. Also, all of the decaying DM with such large decay rates will disappear into dark radiations before the recombination and thus the comoving density of the remaining cold DM is unchanged with respect to $z$, so that different decay rates will start to change $\Phi$ differently, mostly at the very early universe at $z>1000$. Meanwhile, the detailed calculation in Appendix.~\ref{sec:app_B} shows the anti-correlation between $\Gammadcdm$ and $|\Phi|$ in the short-lived case. This anti-correlated feature will be also related to the explanations of the results in Sec.~\ref{sec:growth} and~\ref{sec:kSZ}.

\begin{figure}[htbp!]
\centering
\includegraphics[scale=0.5]{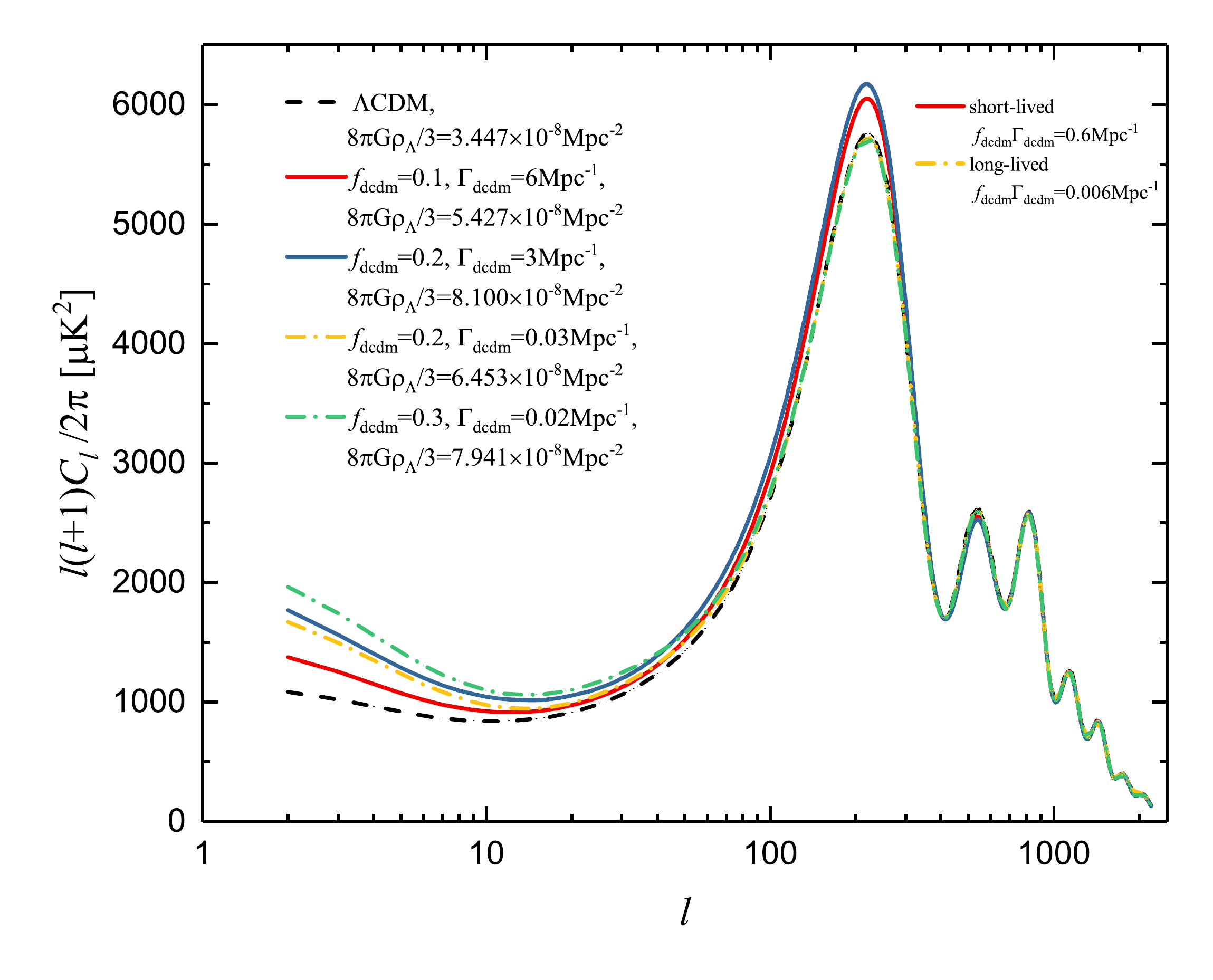}
\caption{\label{fig:CMB_f} Comparison of CMB TT angular power spectrum by varying the abundance $\fdcdm$ and the decay rate $\Gammadcdm$. For a given value $\fdcdm\cdot\Gammadcdm$ (e.g., $\fdcdm\cdot\Gammadcdm=0.006/0.6$ Mpc$^{-1}$ in the long/short-lived case), there exists an underlying degeneracy between $\fdcdm$ and $\Gammadcdm$, especially at the scales of $\ell\gtrsim200$. In either case (short-lived or long-lived), a larger $\fdcdm$ will leave a stronger impact on the low-$\ell$ modes. In the long-lived case, the larger $\fdcdm$ leads to an increase in the spectrum at $\ell\lesssim80$ as the present-day value of $\rho_\Lambda$ is enhanced for keeping the $\theta_{\rm MC}$ fixed same as the one in the $\Lambda$CDM (see details in Appendix~\ref{sec:app_A}). In the short-lived case, increasing $\fdcdm$ will not only boost the low-$\ell$ power but also the height of the first peak.}
\end{figure}

Now let us focus on the effects from the DM density fraction $\fdcdm$. Since the DM density in terms of the Taylor series of $\exp{(-\Gammadcdm t)}$ can be written as $\Omega_c (t) = [1- \fdcdm\Gammadcdm t+ \fdcdm\mathcal{O}(\Gammadcdm t)^2]\Omega_c^{ini}$, the impacts on cosmological observables at the linear level are expected to depend on the quantity $\fdcdm\cdot\Gammadcdm$ only, which means that the cosmological data could not break this degeneracy between $\fdcdm$ and $\Gammadcdm$ when $\Gammadcdm\ll H_0$ and cosmological parameters are fixed. Nevertheless, the condition ($\Gamma_{\rm dcdm}\ll H_0$) can be relaxed for a specific cosmological probe. For example, from Fig.~2(a) we can observe that, changes on TT spectrum at $\ell \gtrsim 100$  by varying $\Gamma_{\rm dcdm}$ from 0.003 to 0.3 Mpc$^{-1}$ with respect to the $\Lambda$CDM one are almost negligible. This negligible effect is also observed at the EE power spectrum in Fig.~1 (Ref.~\cite{Poulin:2016nat}). Thus, the condition can be relaxed to $\Gamma_{\rm dcdm} \ll  H (z\simeq 1100)\approx 5$ Mpc$^{-1}$  for the CMB signals at small scales, since we have fixed the $\theta_{\rm MC}$ (leading to $\rho_{\Lambda}$ adjusted accordingly for a given  $\Gamma_{\rm dcdm}$) and the primary CMB acoustic peaks are then insensitive to the $\rho_{\Lambda}$. Morever, as observed in Fig.~\ref{fig:CMB_f}, when $\fdcdm\cdot\Gammadcdm$ is fixed in either long-lived or short-lived case, the changes in TT spectrum at $\ell\gtrsim 80$ are essentially invisible by varying $\fdcdm$. However, such strong degeneracy is entirely broken at the large scales since in our calculations $\theta_{\rm MC}$ is fixed and thus $\rho_{\Lambda}$ is changed accordingly by varying $\fdcdm$. The short-lived one can have an additional impact on the TT spectrum, boosting the first-peak amplitude, which might be due to contributions from the higher order terms $\mathcal{O}(\Gammadcdm t)^2$.

For the sake of simplicity, here we only discuss the effects in the CMB TT power spectrum. It is straightforward to extend the discussion into CMB polarization patterns (for details of impacts on the polarization see Ref.~\cite{Poulin:2016nat}). However, when performing the joint analysis in Sec.~\ref{sec:results}, all Planck CMB power spectra of TT, TE and EE are included.

\subsection{Impacts of DM decay on the growth of the large-scale structure }\label{sec:growth}

\begin{table}[h!]
  \caption{ \label{table:RSD} The RSD data used in this study. Note that, the effects from data correlation in the likelihood for the MCMC analysis have been taken into account by use of the RSD data covariance matrix  at Tab. 8 in Ref.~\cite{Alam:2016hwk}.} \centering
	\begin{tabular}{ccccc}
		\hline
		\multicolumn{1}{c}{\ \ \ \ \ z\ \ \ \ \ } & \multicolumn{1}{c}{$f\sigma_8(z)$} & \multicolumn{1}{c}{Reference}\ \  \\
		\hline
		0.02	 & 0.360 $\pm$	0.040 & \cite{Hudson:2012gt} \\
		0.067 & 	0.423 $\pm$	0.055 & \cite{Beutler:2012px} \\
		0.10	 & 0.37 $\pm$	0.13 & \cite{Feix:2015dla} \\
		0.17	 & 0.51 $\pm$	0.06  & \cite{Song:2008qt} \\
		0.22	 & 0.42 $\pm$	0.07 & \cite{Blake:2011rj} \\
        0.38     & 0.497$\pm$   0.045 & \cite{Alam:2016hwk} \\
		0.41	 & 0.45 $\pm$	0.04 & \cite{Blake:2011rj} \\
		0.51	 & 0.458 $\pm$	0.038 & \cite{Alam:2016hwk} \\
		0.6	 & 0.43 $\pm$	0.04 & \cite{Blake:2011rj} \\
		0.61     & 0.436 $\pm$  0.034 & \cite{Alam:2016hwk} \\
		0.77	 & 0.490 $\pm$	0.180 & \cite{Song:2008qt} \\
		0.78	 & 0.38 $\pm$	0.04 & \cite{Blake:2011rj} \\
		0.80	 & 0.47 $\pm$	0.08 & \cite{delaTorre:2013rpa} \\
		\hline
	\end{tabular}
\end{table}

\begin{figure}
	\centering
	\includegraphics[scale=0.4]{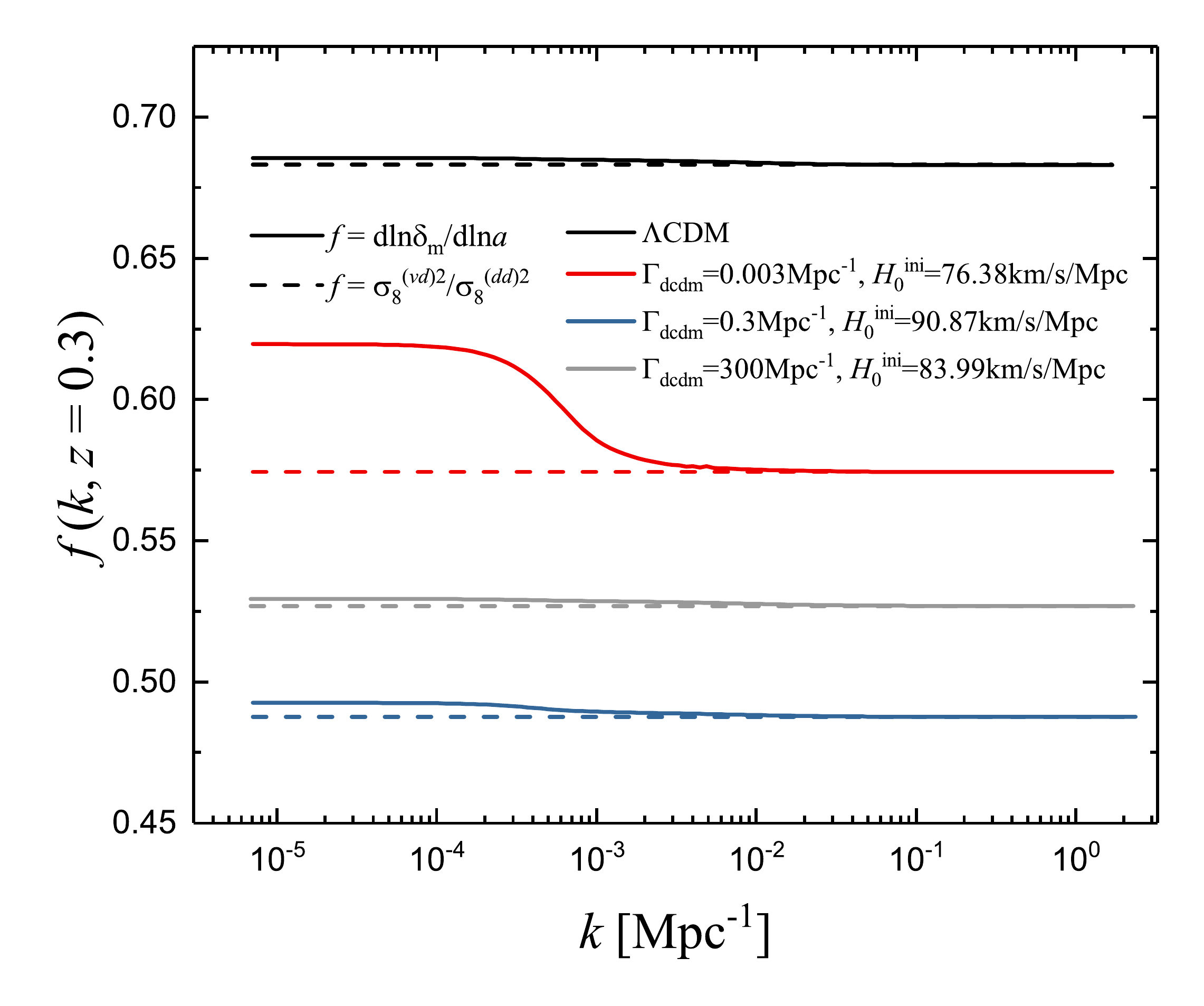}
	\caption{\label{fig:growth_rate} Comparison of the matter growth rate at $z=0.3$. For the each model, the growth rates $f$ measured from $d\ln\delta_{\rm m}/d\ln a$ and $\left(\sigma_8^{vd}\right)^2/\left(\sigma_8^{dd}\right)^2$ are almost identical at the scales of $k\gtrsim0.01$ Mpc$^{-1}$. The decay-induced suppression in $f$ is apparent and a larger decay rate will lead to a stronger suppression.}
\end{figure}

A measurement of the large-scale structure growth may give further constraints on the DCDM model. In this study, we focus on the investigation of decay effects in the RSD and provide a constraint on the model by using the $f\sigma_8$ data (listed in Table \ref{table:RSD}) from~\cite{Costa:2016tpb}. The model independent matter growth rate is defined as 
\begin{equation}\label{eq:define_f}
f \equiv \frac{dln\delta_m}{dlna} = \frac{\mathcal{H}^{-1}}{\delta_m}\delta_m'\,,
\end{equation}
which in generally is a function of the redshift $z$, however the growth rate might be not sensitive to the small scales of $k\gtrsim0.01$ Mpc$^{-1}$. In the DCDM model, the overdensity of total matter $\delta_m$ is modified as 
\begin{equation}\label{eq:define_deltam}
 \delta_m \equiv \frac{\rho_{\rm sdm}\delta_{\rm sdm}+\rho_{\rm dcdm}\delta_{\rm dcdm}+\rho_b\delta_b}{\rho_m}\,,
\end{equation}
where $\rho_m = \rho_{\rm sdm} + \rho_{\rm dcdm} + \rho_b$.

We usually use the variance of the density field smoothed within $8h^{-1}$ Mpc (denoted as $\sigma_8^{dd}$ or $\sigma_8$) and the velocity-density correlation with the same smoothing scale $\sigma_8^{vd}$, to estimate the growth rate through

\begin{equation}\label{eq:define_f1}
f \equiv [\sigma_8^{vd}(z)]^2/[\sigma_8^{dd}(z)]^2\,.
\end{equation}

There is no modification in the formalism of $\delta_{\rm  dcdm}'$ from the DCDM model in the synchronous gauge (see Eq.~\ref{eq:ContinuityDecay}), so we have
\begin{equation}\label{eq:deltam_continuity}
\delta_m' = -\theta_m - \frac{h'}{2}\,,
\end{equation}
which is completely different from interacting dark matter-dark energy scenarios~\cite{Costa:2016tpb,Marcondes:2016reb,An:2018vzw,Costa:2019uvk}, and thus the measurements of $f$ from Eqs.~\ref{eq:define_f} and~\ref{eq:define_f1} should be almost identical at small scales, confirmed by the illustration in Fig.~\ref{fig:growth_rate}.



For $f\sigma_8$ data, the measurements are based on the coherent motion of galaxies that characterized by the continuity equation, which reads 
\begin{equation}\label{eq:continuity_G}
\theta_G = -\mathcal{H} \beta \delta_G - \frac{h'}{2}\,.
\end{equation}
Here we have introduced the RSD parameter $\beta$, with $\beta\equiv f/b$, where $b$ is the galaxy bias ($b = \delta_G/\delta_m$). Under the assumption of $\theta_G = \theta_m$, we can therefore straightly obtain the continuity equation for the total matter by combining Eq.~\ref{eq:define_f} and Eq.~\ref{eq:deltam_continuity} as 
\begin{equation}\label{eq:thetam}
\theta_m = -\mathcal{H} f \delta_m - \frac{h'}{2}\,,
\end{equation}
which indicates the strong dependence of the divergence of the velocity field $\theta_m$ on the growth rate.

\begin{figure}[htbp!]
	\subfloat[]{
		\includegraphics[width=0.49\textwidth]{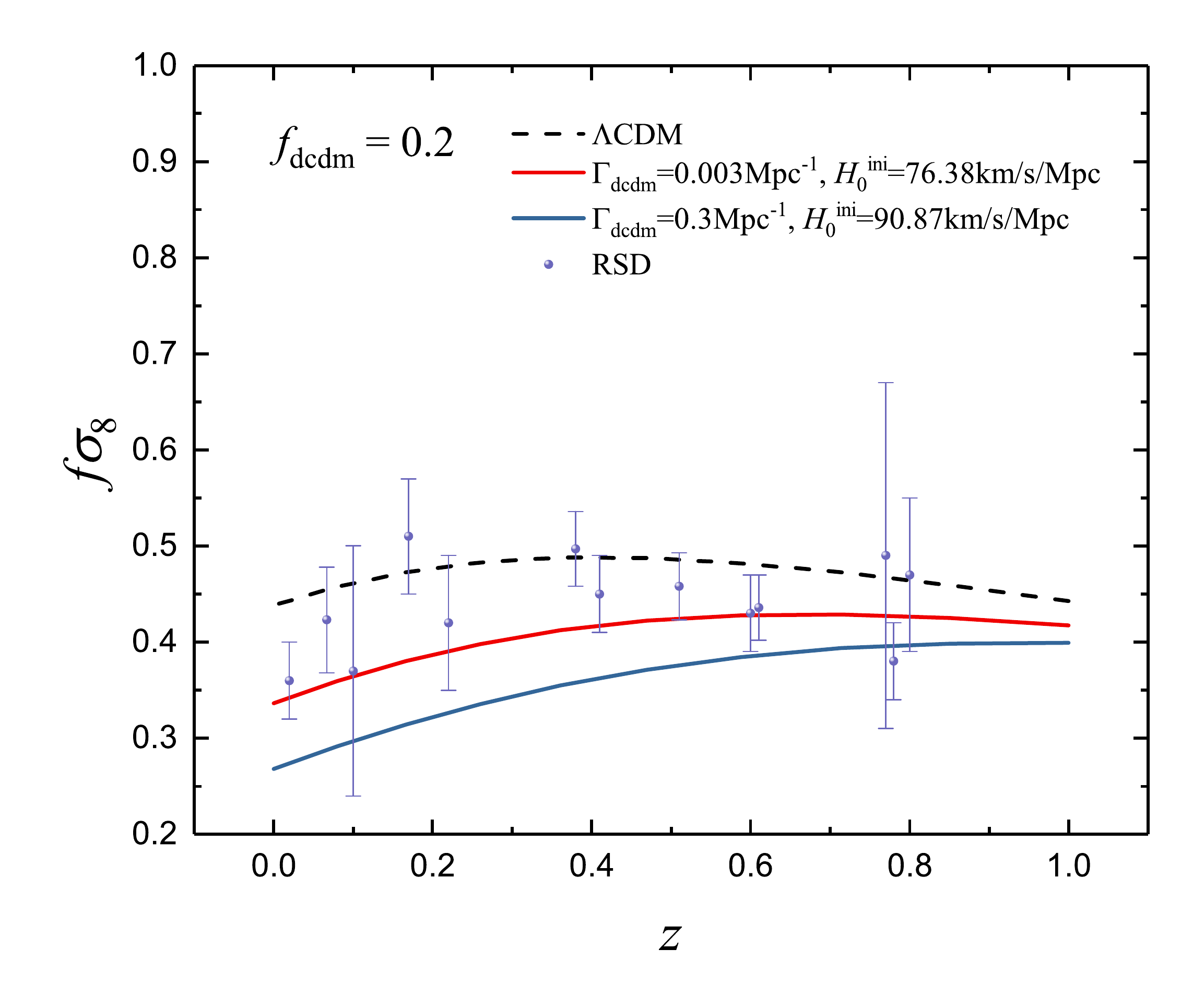}\label{fig:fs8_1}
	}
	\subfloat[]{
		\includegraphics[width=0.49\textwidth]{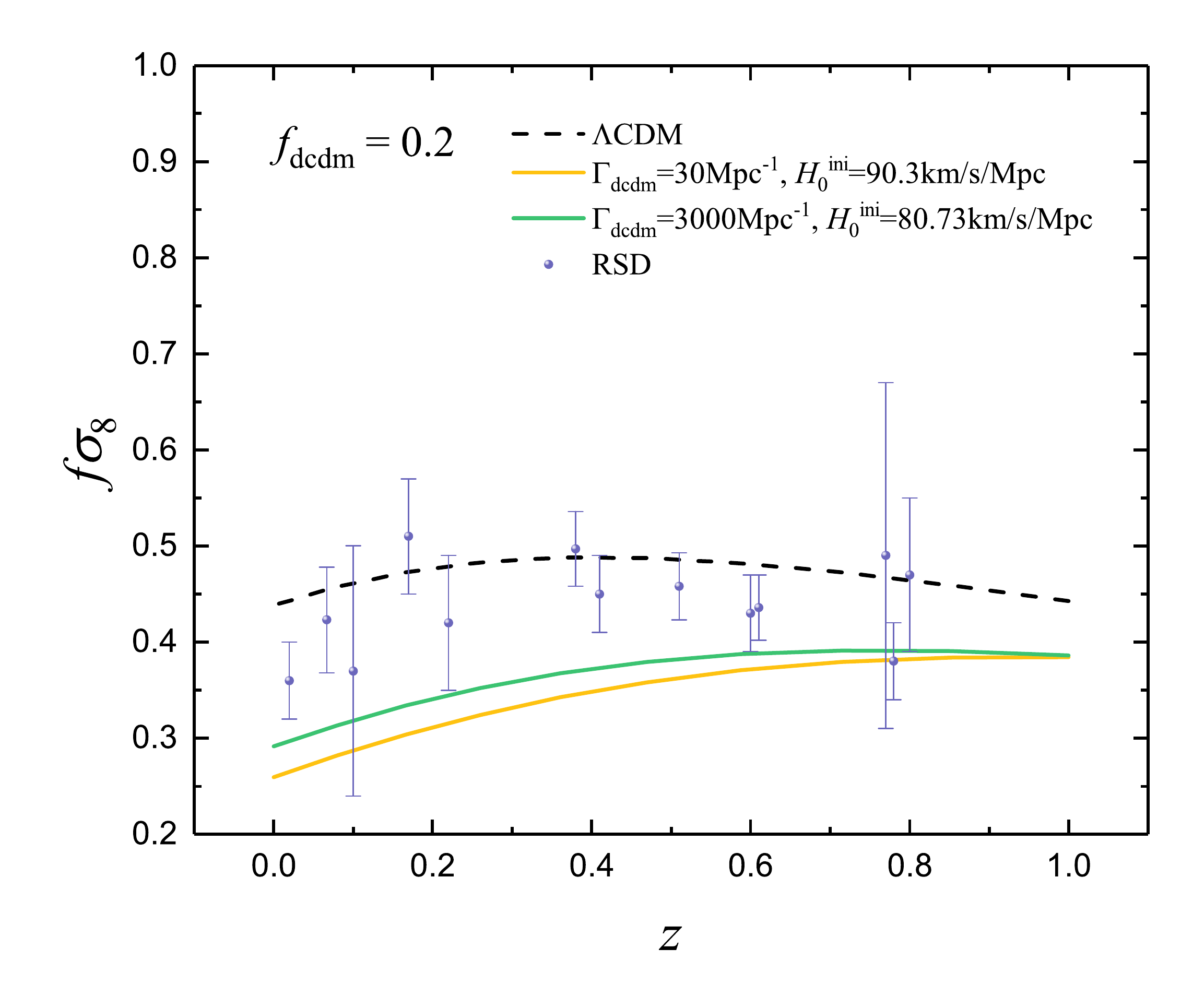}\label{fig:fs8_2}
	}
	\caption{\label{fig:fs8}  Dependence of the evolution of $f\sigma_8$ on several decaying rates, in comparison with the observational data. The decaying DM leads a slightly lower $f\sigma_8$ than the prediction from the $\Lambda$CDM model.}
\end{figure}

We now investigate the evolution of $f\sigma_8(z)$ in the presence of the decaying DM. By varying the decay rate, Fig.~\ref{fig:fs8} shows the predicted $f\sigma_8(z)$, together with the observational data. As seen, for long-lived DCDM $f\sigma_8(z)$ will decrease as $1/\Gammadcdm$ decreases monotonically, whereas for short-lived DCDM the anti-correlated feature (mentioned in Sec.~\ref{sec:CMB}) still exists, which is consistent with the impact of the decay on $\Phi$ shown in Fig.~\ref{fig:bf_phi} (see further discussion later).

Moreover, one may invoke the DCDM model, since the prediction on $f\sigma_8(z)$ from $\Lambda$CDM is slightly higher than the observed data\footnote{For example, the $\chi^2$ of $f\sigma_8$ for $\Lambda$CDM and the DCDM model with $\Gammadcdm=0.003$ Mpc$^{-1}$ (the red line in Fig.~\ref{fig:fs8_1}) are 28.07 and 11.51, respectively, implying the DCDM model have a better fit to observational growth rate data points with respect to the $\Lambda$CDM one.} and such overestimate can be reduced by allowing for our DCDM model that suppresses the growth rate of cosmological perturbations. Also, the $f\sigma_8(z)$ in the low-z region ($z\lesssim1$) is very sensitive to the value of $\Gammadcdm$, and thus we expect that  the current and the future $f\sigma_8(z)$ data combined with the Planck CMB observations will provide a tight constraint on the decaying model.

\subsection{Impacts of DM decay on baryon velocity field and kSZ effect}\label{sec:kSZ}

The baryon velocity field is an accurate tracer of the cosmic evolution that is sensitive to the DM decay, and thus it may pose a strong constraint for the DCDM model. 

In the linear perturbation theory, the evolution of the baryon peculiar velocity in the Fourier space follows
\begin{equation}\label{eq:vb}
v_{\mathrm{b}}' = -\mathcal{H}v_{\rm b} + k\psi\,,
\end{equation}
where the subscript ``b'' represents baryon. The corresponding statistical information on $v_{\mathrm{b}}$ can be obtained by translating $v_{\mathrm{b}}$ into the root-mean-square velocity dispersion of baryon
\begin{equation}\label{eq:bf}
\left< v_{\mathrm{b}}^2 \right> = \int d^3k W_r^2(k) P_v(k)\,
\end{equation}
within a sphere of radius $r$, where $P_v(k)$ is the power spectrum of $v_{\rm b}$, $W_r(k)$ is the top-hat filter in the $k$-space, and $\langle v_{\mathrm{b}}^2 \rangle^{1/2}$ is the bulk flow of baryon on such radius $r$.

\begin{figure}[tb]
	\subfloat[]{
		\includegraphics[width=0.49\textwidth]{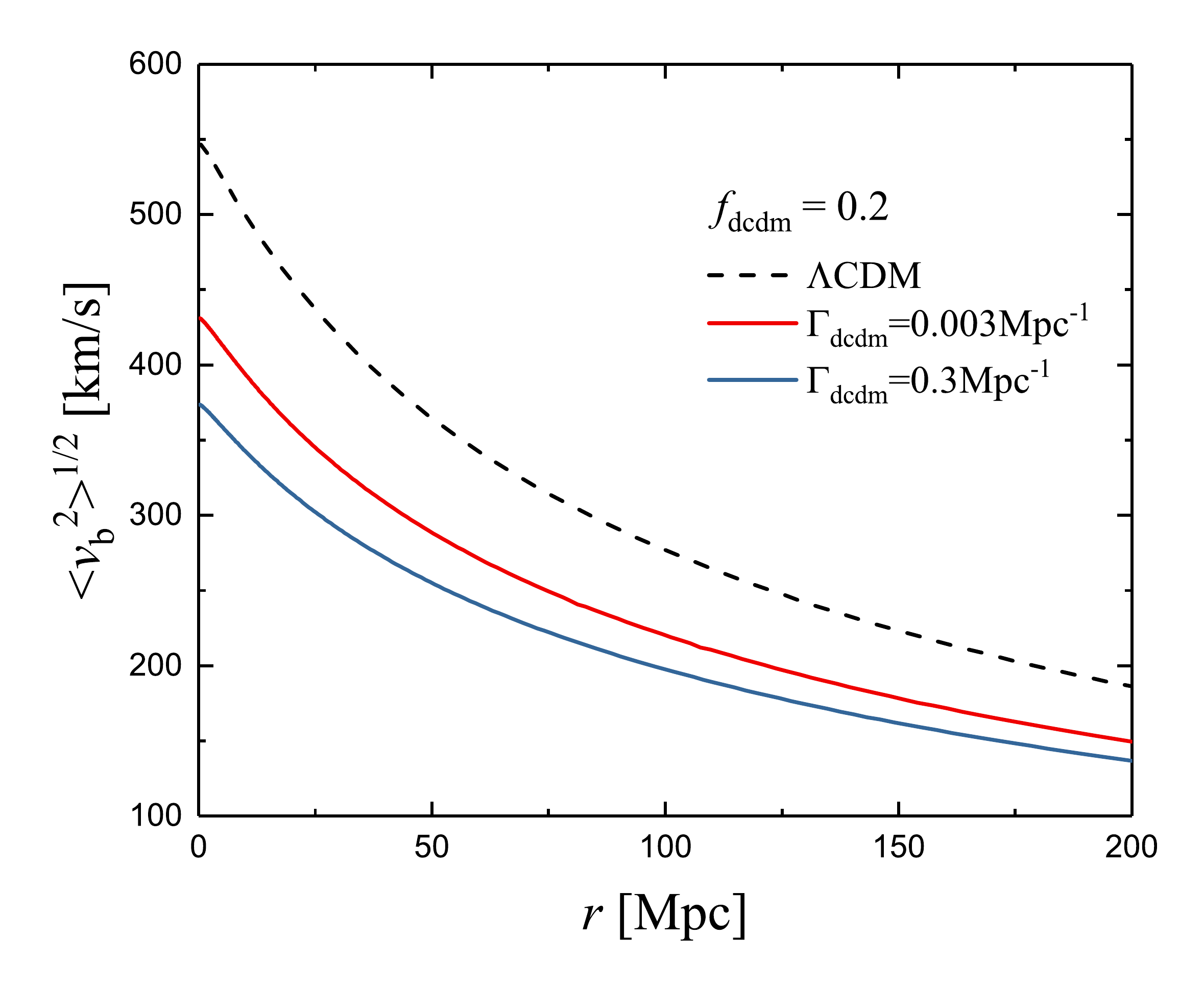}\label{fig:bf_1}
	}
	\subfloat[]{
		\includegraphics[width=0.49\textwidth]{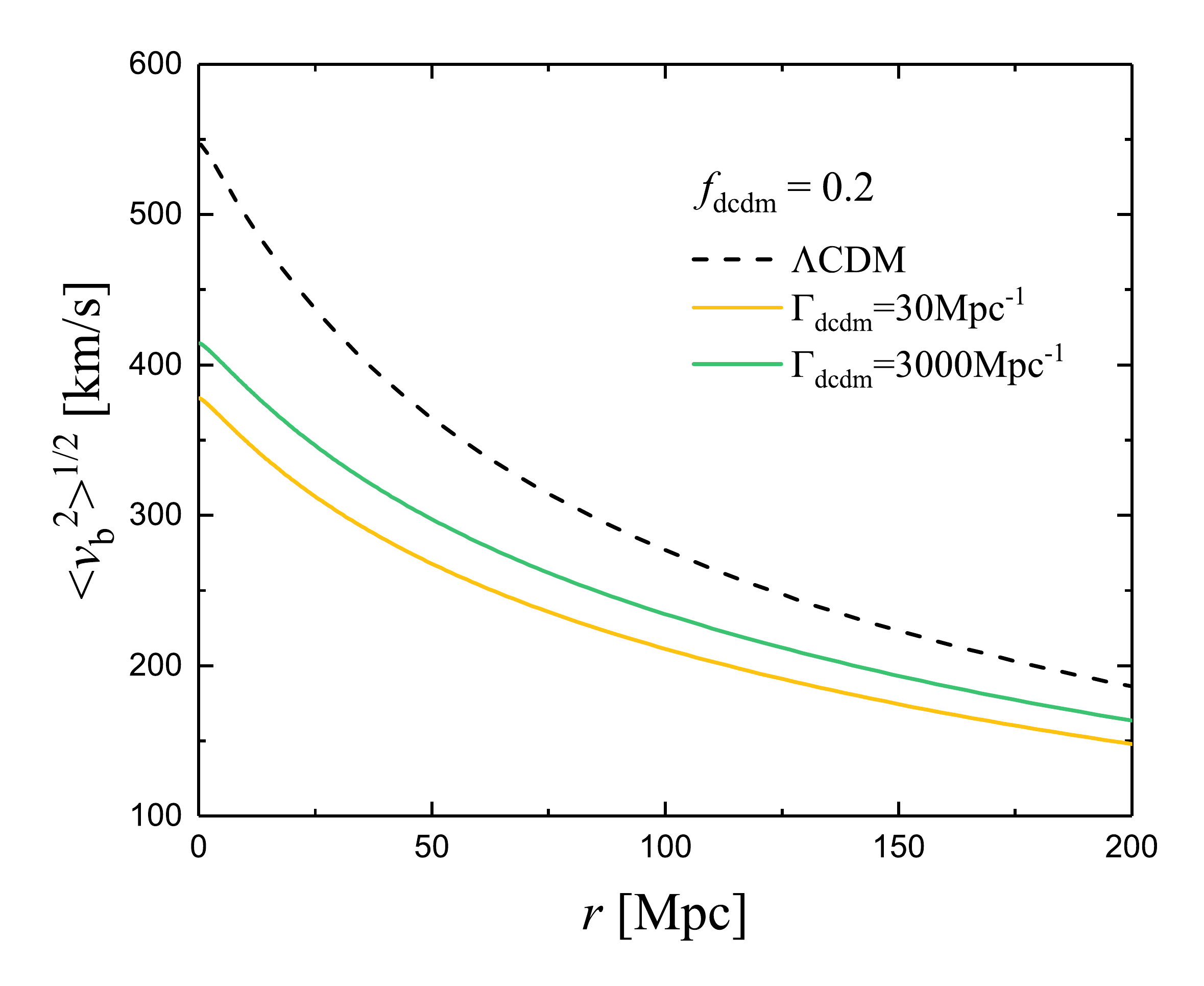}\label{fig:bf_2}
	}
	\caption{\label{fig:bf} Impacts of the decaying DM with several decay rates on the baryon velocity field $v_{\rm b}$ at $z=0$, in the long-lived (left) and short-lived (right) cases.}
\end{figure}

\begin{figure}[tb]
	\subfloat[]{
		\includegraphics[width=0.49\textwidth]{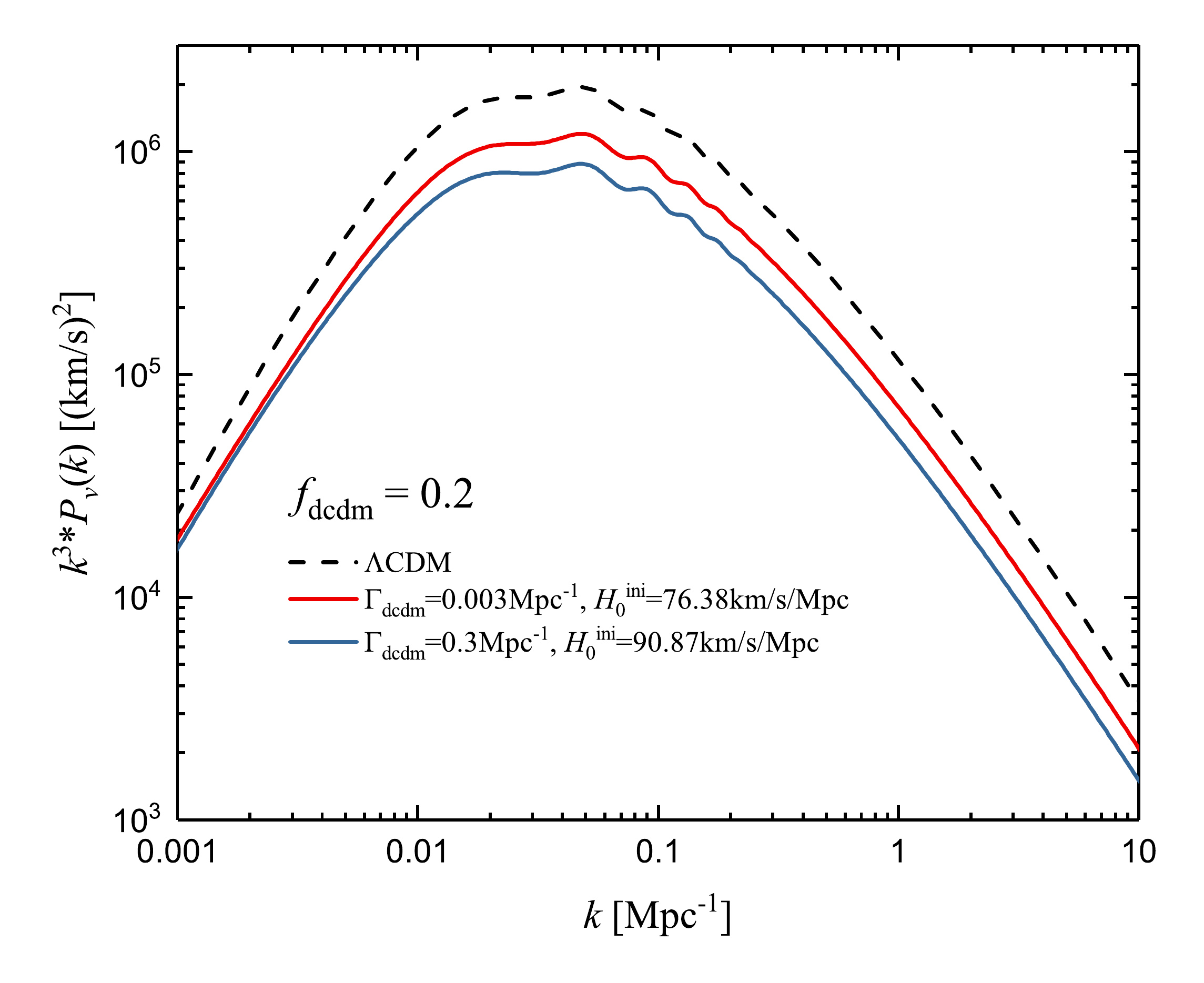}\label{fig:Pv_1}
	}
	\subfloat[]{
		\includegraphics[width=0.49\textwidth]{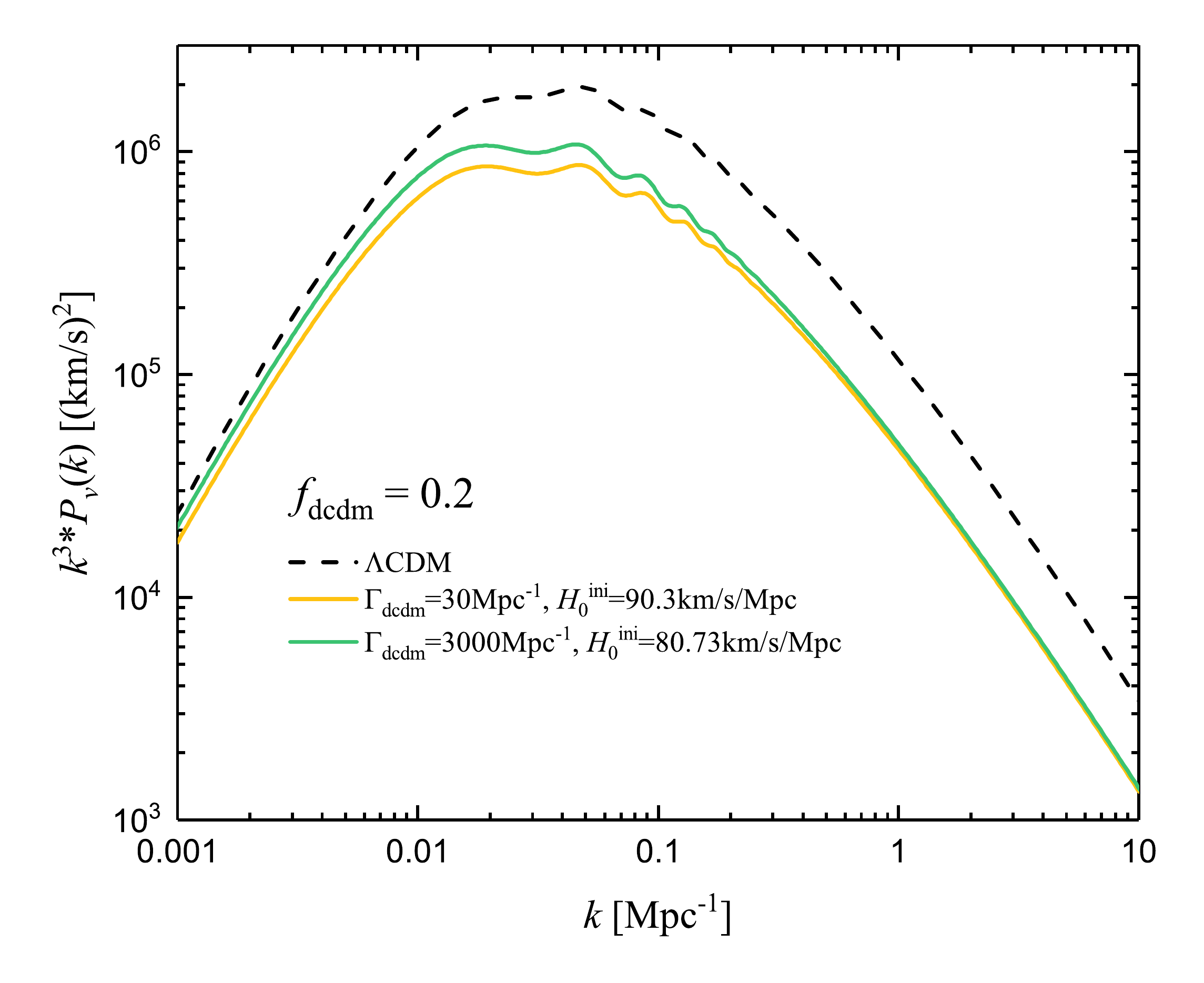}\label{fig:Pv_2}
	}
	\caption{\label{fig:bf_Pv} An illustraion of $k^3P_v$ as a function of $k$ at $z=0$ for $\Lambda$CDM and DCDM model under parameter convention in Fig. \ref{fig:bf}.}
\end{figure}

\begin{figure}[tb]
	\subfloat[]{
		\includegraphics[width=0.49\textwidth]{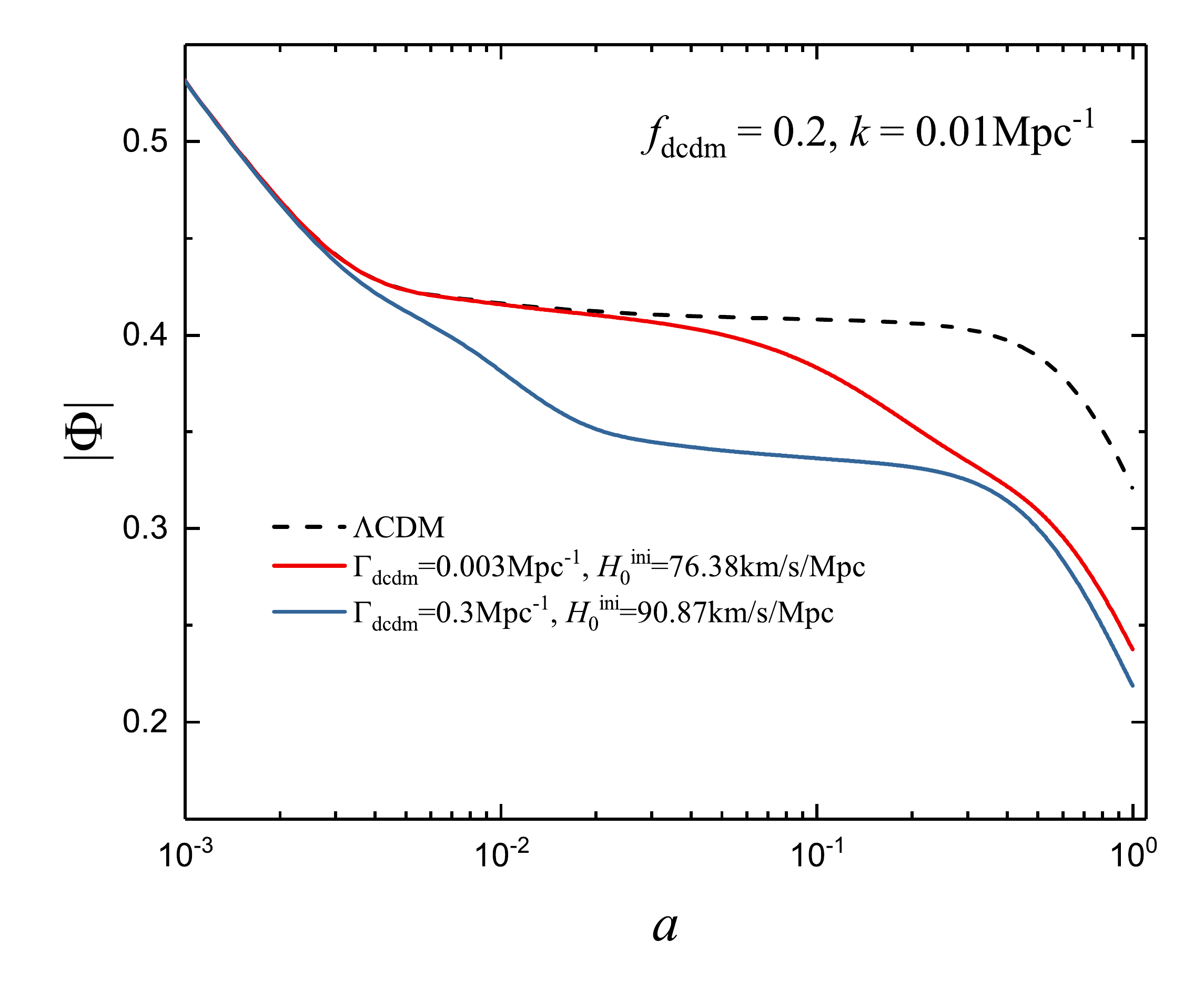}\label{fig:phi1_k0.01}
	}
	\subfloat[]{
		\includegraphics[width=0.49\textwidth]{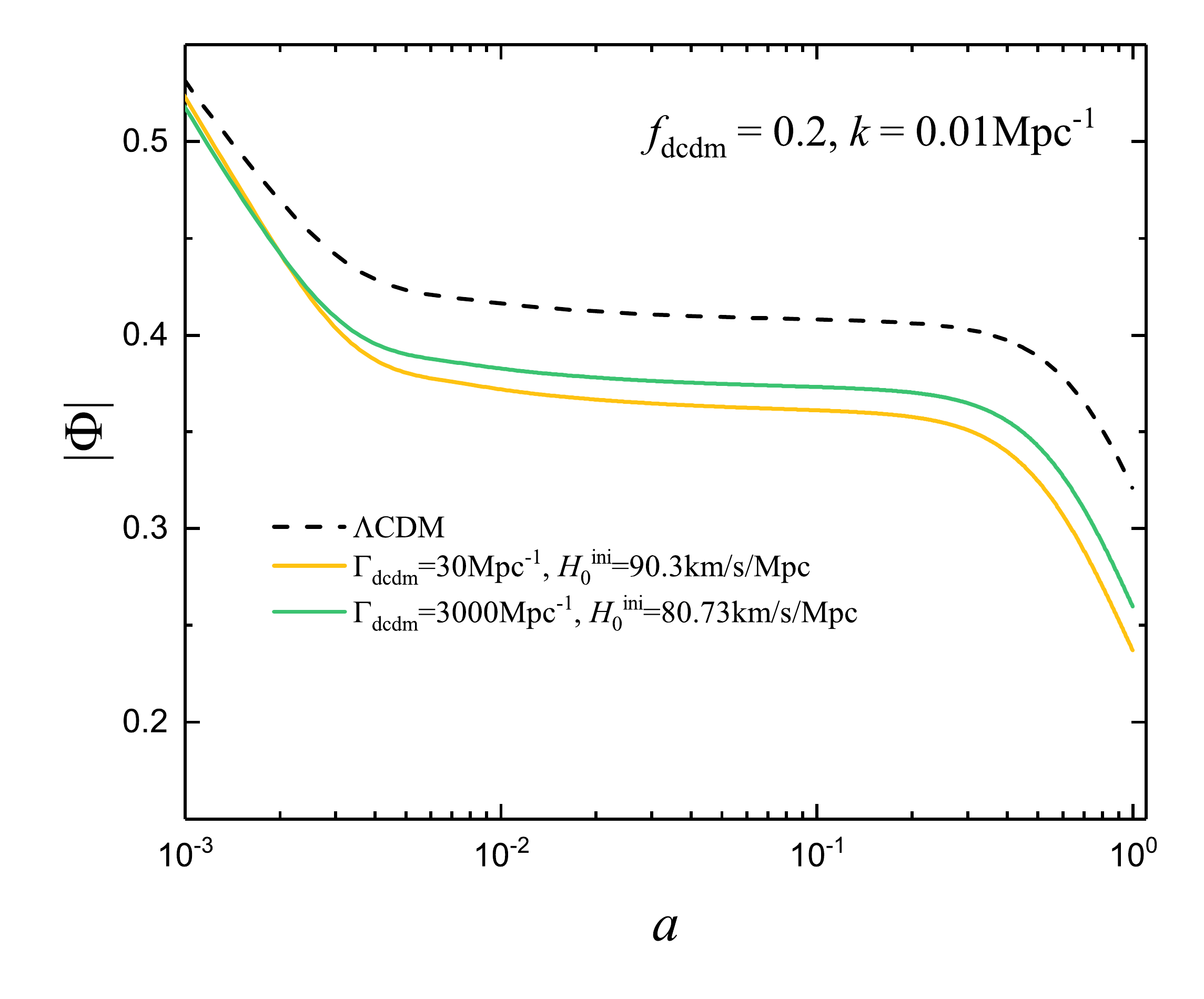}\label{fig:phi2_k0.01}
	}
	\caption{\label{fig:bf_phi} Dependence of the evolution of $|\Phi|$ on $\Gammadcdm$ at the wavenumber of $k=0.01$ Mpc$^{-1}$. The calculations are based on the same initial condition as in Fig. \ref{fig:fs8} \& \ref{fig:bf} and the behaviours of $|\Phi|$ reveal the features of the baryon velocity field shown in those figures.}
\end{figure}

Fig.~\ref{fig:bf} illustrates a strong suppression in the baryon velocity field across all scales.  A larger decay rate will lead to a stronger suppression for the long-lived DM model with $\Gammadcdm\lesssim 0.3$ Mpc$^{-1}$. However, the short-lived DM impacts on the velocity field in the other way around and a larger decay rate will weaken the suppression from Fig.~\ref{fig:bf_2}. The trend of the suppression can be clearly understood as follows: 1) the top-hat filter in Eq.~\ref{eq:bf} results in the contribution of $P_v(k)$ to $\left< v_{\mathrm{b}}^2 \right>$ almost from the modes with $k\lesssim 1$ Mpc$^{-1}$; 2) $k^3P_v(k)$ on very large scale is insignificant, and what really makes sense is the statistical information within $k$ range we plot in Fig.~\ref{fig:bf_Pv}; 3) $P_v(k)$ is the auto-correlation of $v_{\mathrm{b}}$, and on small scales the baryon velocity field $v_{\rm b}$ is proportional to the overdensity $\delta_b$ that is still controlled by the gravitational potential. The deeper the potential is, the stronger the resulting growth is, which is demonstrated by varying the decay rates shown in Fig.~\ref{fig:bf_phi}. 

Furthermore, we also expect that the decaying DM would affect the velocity field of the universe even before the reionization, which might be able to be tested by the kinetic Sunyaev-Zel'dovich effect (kSZ) that directly encode the information of the peculiar velocity field of the matter distribution and the baryonic matter density. The kSZ induced temperature anisotropies in a given direction ($\hat{\bm n}$) on the sky is given by~\cite{Sunyaev:1972eq} 

\begin{equation}
\frac{\Delta T(\hat{\bm n})}{T_{\mathrm{CMB}}} =
-\int_{t_{\mathrm{*}}}^{t_0} n_{\mathrm{e}}\sigma_Te^{-\tau}(\bm{v}\cdot\hat{\bm n})\mathrm{d}t\,,
\end{equation}
where $n_{\rm e}$ is the electron density, $\sigma_T$ the Thomson cross-section, $\tau$ the optical depth, $\bm{v}$ the peculiar velocity vector of ionized electrons and $dt$ corresponds to the distance along the line-of-sight (l.o.s) from $t_{\mathrm{*}}$, which will happen before the reionization as we want to collect all the kSZ signals along l.o.s, to the present day $t_0$. Assuming $n_{\rm e}(x,z)=\chi_{\rm e}\bar{n}_{\rm e}(0)a^{-3} [1+\delta_{\mathrm{e}}(x,z)]$, where $\chi_{\rm e}$ is the ionization fraction, $\bar{n}_{\rm e}(0)$ the mean electron density today and $x$ is the position in comoving coordinates. For simplicity, in this study we only consider the contribution from homogenous kSZ, not from the patchy effect. The ``patchy'' reionization kSZ in the fiduical model is subdominant with respect to the homogeneous one. Also, the patchy effect is stronly model dependent and depends on $\chi_{\mathrm{e}}$. Moreover, one could separate the late-time and reionization kSZ by considering a higher-order statistic proposed in Ref.~\cite{Smith:2016lnt}. With this concern, in this study $\chi_{\mathrm{e}}$ is assumed to be a function of the redshift $z$ only, and the contribution from inhomogeneous $\chi_{\mathrm{e}}$ is neglected. The kSZ-induced temperature anisotropies then can be computed as 

\begin{equation}
\frac{\Delta T(\hat{\bm n})}{T_{\rm CMB}} =
\bar{n}_{\rm e}(0) \sigma_T \int a^{-2} \chi_{\rm e}(z) e^{-\tau}(\bm{p}\cdot\hat{\bm n})\mathrm{d}x\,, 
\end{equation}
in which $\bm{p}\equiv (1+\delta_{\mathrm{e}})\bm{v}$ is the peculiar momentum of the free electrons and can be decomposed into the gradient part $\bm{p}_{\rm E}$ and the curl part $\bm{p}_{\rm B}$. As known, there is no contribution from  $\bm{p}_{\rm E}$ to the kSZ effect~\cite{Vishniac:1987wm,Zhang:2003nr} since the gradient part cancels out when integrating along the line of sight. Thus we only consider the curl part and the corresponding power spectrum is computed by 

\begin{equation}\label{eq:pB}\begin{split}
P_{\mathrm{B}}(k) = \langle\bm{p}_{\mathrm{B}}^*(\bm{k}) \cdot \bm{p}_{\mathrm{B}}(\tilde{\bm{k}})\rangle =
&\int\frac{\mathrm{d}^3\bm{k}'}{(2\pi)^3}
\int\frac{\mathrm{d}^3\tilde{\bm{k}}'}{(2\pi)^3}
\langle \delta_{\mathrm{e}}^*(\bm{k}-\bm{k}')\delta_{\mathrm{e}}(\tilde{\bm{k}}-\tilde{\bm{k}}')
v^*(\bm{k}') v(\tilde{\bm{k}}') \rangle \\
&\times \lvert \bm{k}' \rvert \lvert \tilde{\bm{k}}'
\rvert \beta(\bm{k},\bm{k}') \cdot \beta(\tilde{\bm{k}},\tilde{\bm{k}}')\,,
\end{split}
\end{equation}
where $P_{\rm B}(k)$ is the power spectrum of $\bm{p}_{\rm B}$ and the kernel $\beta(\bm{k},\bm{k}') = [\bm{k}'-\bm{k}(\bm{k}\cdot\bm{k}')/\bm{k}^2]/\bm{k}'^2$. In the linear perturbation theory and for $k\eta\gg1$, the peculiar velocity $\bm{v}$ is related to the overdensity $\delta_e$ through
\begin{equation}
v=-\delta_{\mathrm{e}}'/k=-aHf_{\mathrm{e}}(a)\delta_{\mathrm{e}}/k\,,
\end{equation}
in which $f_{\mathrm{e}}=\frac{d\ln \delta_{\mathrm{e}}}{d\ln a}$ is the growth factor for the electron overdensity that is same as the baryonic one at the linear level. With $aHf_{\mathrm{e}}=a\dot{D_{\mathrm{e}}}/{D_{\mathrm{e}}}$ by setting $D_{\mathrm{e}}(z) \equiv \delta_{\mathrm{e}}(z)/\delta_{\mathrm{e}}(0)$, Eq.~\ref{eq:pB} can be re-expressed as
\begin{equation}\label{eq:pB_e}\begin{split}
P_{\mathrm{B}}(k,z) =
& \frac{a^2}{2} \int \frac{\mathrm{d}^3\bm{k}'}{(2\pi)^3}
\bigg(\frac{\dot{D_{\mathrm{e}}}}{D_{\mathrm{e}}}\bigg)^2 P(k',z) P(k-k',z)\\
&\times [W_g(k-k')\beta(\bm{k},\bm{k}') + W_g(k')\beta(\bm{k},\bm{k}-\bm{k}')]^2\,,
\end{split}
\end{equation}
where $P(k)$ is the linear power spectrum of the baryon, $W_g(k)$ is the transfer function that suppresses the $P(k)$ at small scales~\cite{Fang:1993hi} and we take it as unity in our calculations. Since the non-linear evolution of matter overdensity would enhance the kSZ signal, this effect can be modeled by a transfer function $T_{NL}$ such that the non-linear power spectrum $P^{NL}(k) \equiv P(k)T_{NL}^2(k)$. Refering to Ref.~\cite{Hu:1999vq,Ma:2001xr,Shaw:2011sy}, one can recast Eq.~\ref{eq:pB_e} for the non-linear effect as  
\begin{equation}\label{eq:pB_e_NL}\begin{split}
P_{\mathrm{B}}(k,z) =
& \frac{a^2}{2} \int \frac{\mathrm{d}^3\bm{k}'}{(2\pi)^3}
\bigg(\frac{\dot{D_{\mathrm{e}}}}{D_{\mathrm{e}}}\bigg)^2 P(k',z) P(k-k',z)\\
&\times [W_g(k-k')T_{NL}(k-k')\beta(\bm{k},\bm{k}')
+ W_g(k')T_{NL}(k')\beta(\bm{k},\bm{k}-\bm{k}')]^2\,.
\end{split}
\end{equation}
In non-standard cosmology, $T_{NL}(k)$ can be accurately fitted from numerical simulations. Several works dedicated to non-linear clustering~\cite{Poulin:2016nat,Enqvist:2015ara} find that the standard Halofit model would significantly overestimate the small-scale power spectrum in the presence of the DCDM models~\cite{Smith:2002dz,Takahashi:2012em}. Recently, Ref.~\cite{Enqvist:2015ara} provides a more accurate fitting formula from N-body simulations to correct such overestimation. However, this fitting formula is guaranteed to be accurate only for the extreme long-lived DCDM model of $\Gamma_{\rm dcdm} \leq 1/31~$Gyr$^{-1}~(i.e., \simeq 1.05\times10^{-4}~$Mpc$^{-1})$, which has a much longer lifetime than the typical values of $\Gamma_{\rm dcdm}$ investigated in this study. With a conservative assumption that the predicted value from this fitting formula for a larger $\Gamma_{\rm dcdm}$ would not be deviated considerably, we therefore calculate the non-linear growth of the baryon via the fitting formula evaluated by a fixed $\Gamma_{\rm dcdm}$ of $1/31$~Gyr$^{-1}$. The kSZ signal thus can be calculated through 
\begin{equation}\label{eq:cl_kSZ}
C_\ell^{kSZ}=T^2_{\rm CMB}\frac{16\pi^2}{(2l+1)^3} (\bar{n}_{\mathrm{e}}(0)\sigma_T)^2
\int^{z_*}_0 (1+z)^4 \chi^2_{\mathrm{e}} \frac{1}{2}\Delta^2_{\mathrm{B}}(k,z)\rvert_{k=l/x}
e^{-2\tau} x(z) \frac{dx(z)}{dz}dz\,,
\end{equation}
where $\Delta_{\mathrm{B}}^2(k,z) = \frac{k^3}{2\pi^2}P_{\mathrm{B}}(k,z)$. The contribution to kSZ signal from high redshifts is quite small since $\chi_{e}\simeq0$ in the fiducial ``sudden'' reionization model used in CAMB~\cite{Lewis:2008wr} with $z_{\rm reio}=9.9$ and $\Delta_z=1.5$, where $z_{\rm reio}$ is the redshift when the reionization fraction is about half of its maximum and $\Delta_z$ is the width of the reionization transition. From numerical tests, we choose the upper limit of the integral as $z_*=12$ as any contributions from the higher ones can be safely neglected.

\begin{figure}[tb]
\centering
\subfloat[]{
	\includegraphics[scale=0.32]{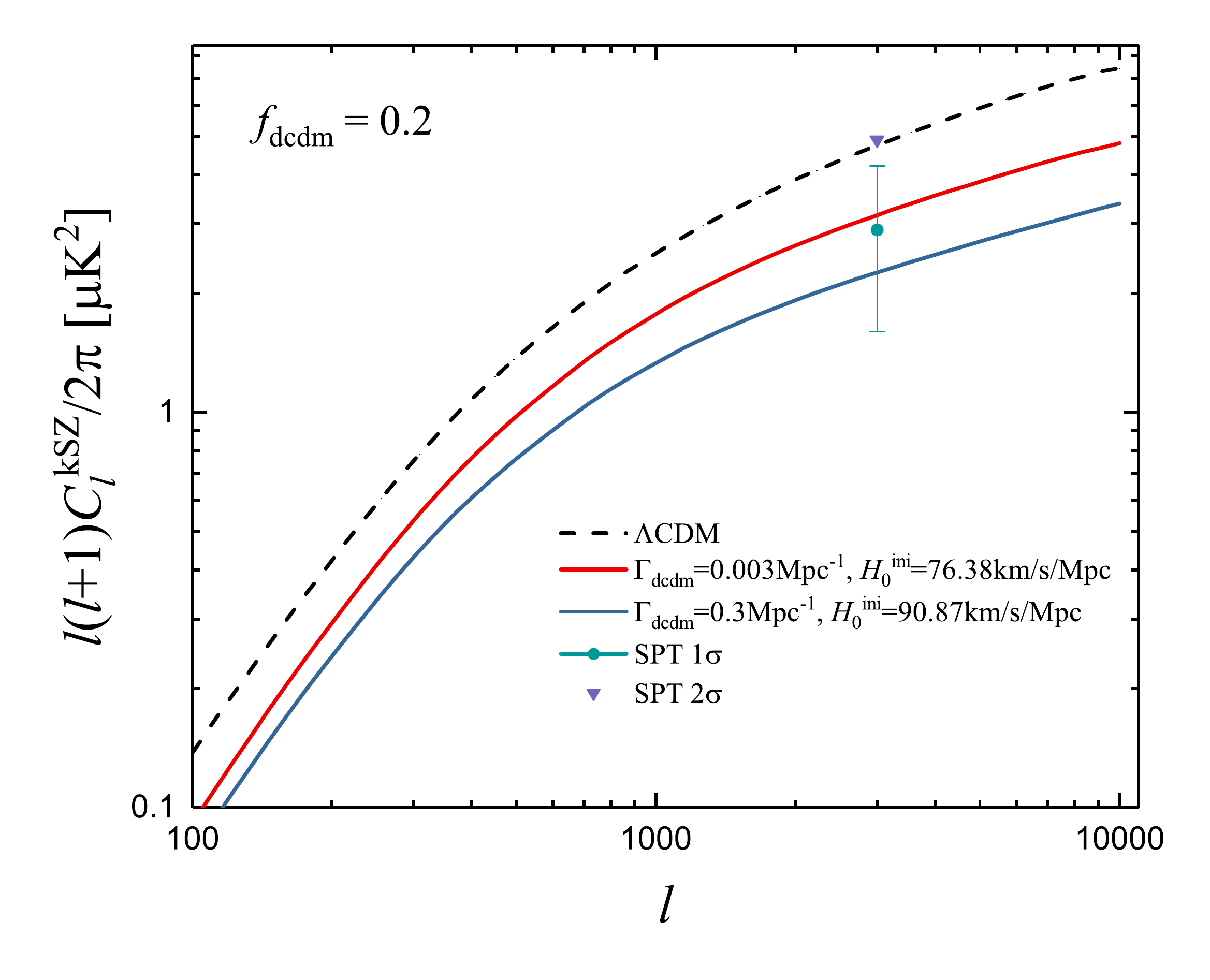}\label{fig:ClkSZ_1}}
\subfloat[]{
	\includegraphics[scale=0.32]{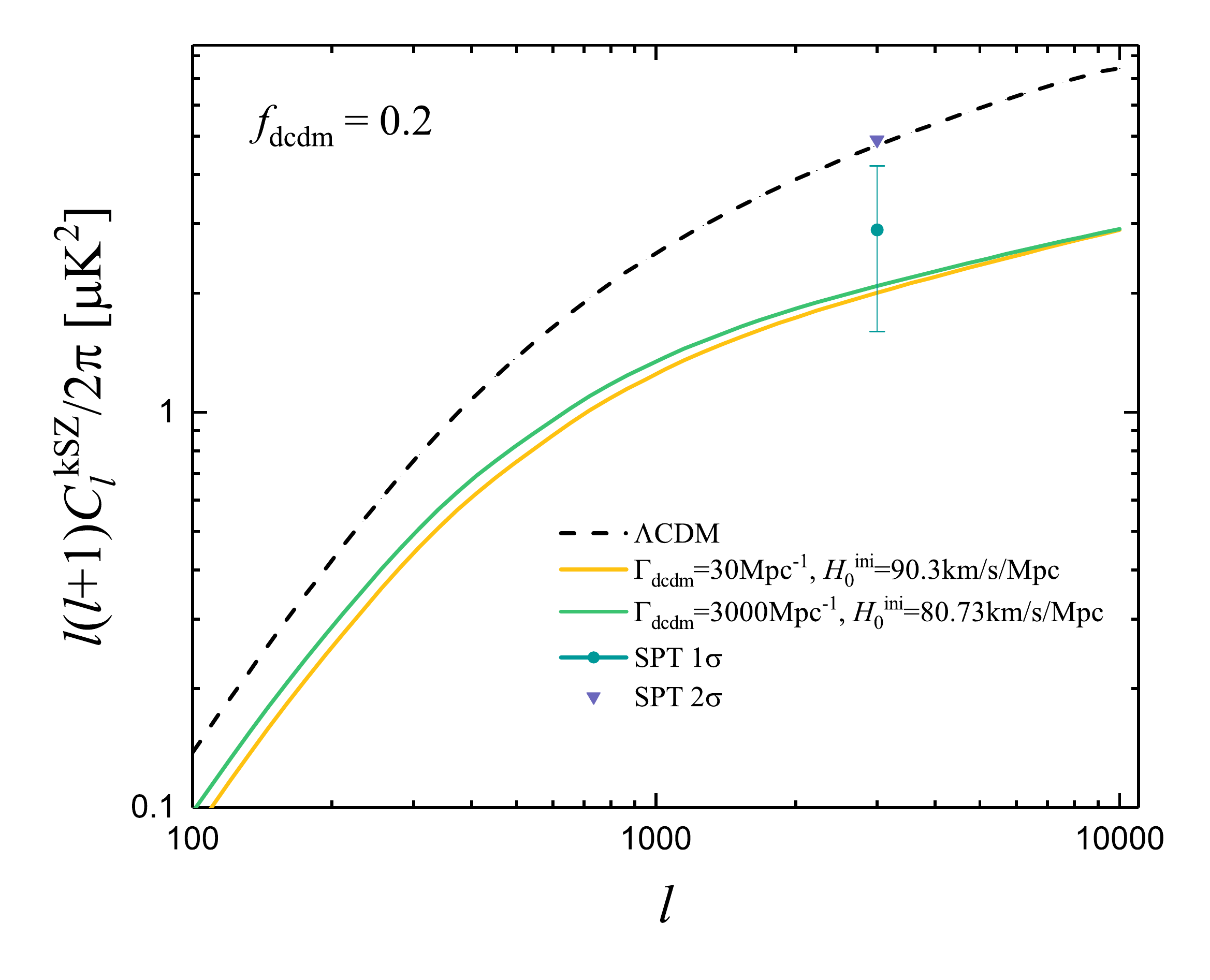}\label{fig:ClkSZ_2}}
\caption{\label{fig:ClkSZ} Comparison of the kSZ power spectrum in the long-lived (left panel) and short-lived (right panel) cases. The dependence of $C_\ell^{kSZ}$ on $\Gammadcdm$ is consistent with the behaviours of the bulk flow as shown in Fig.~\ref{fig:bf}.}

\end{figure}
\begin{figure}[htbp!]
\centering
\subfloat[]{
\includegraphics[scale=0.32]{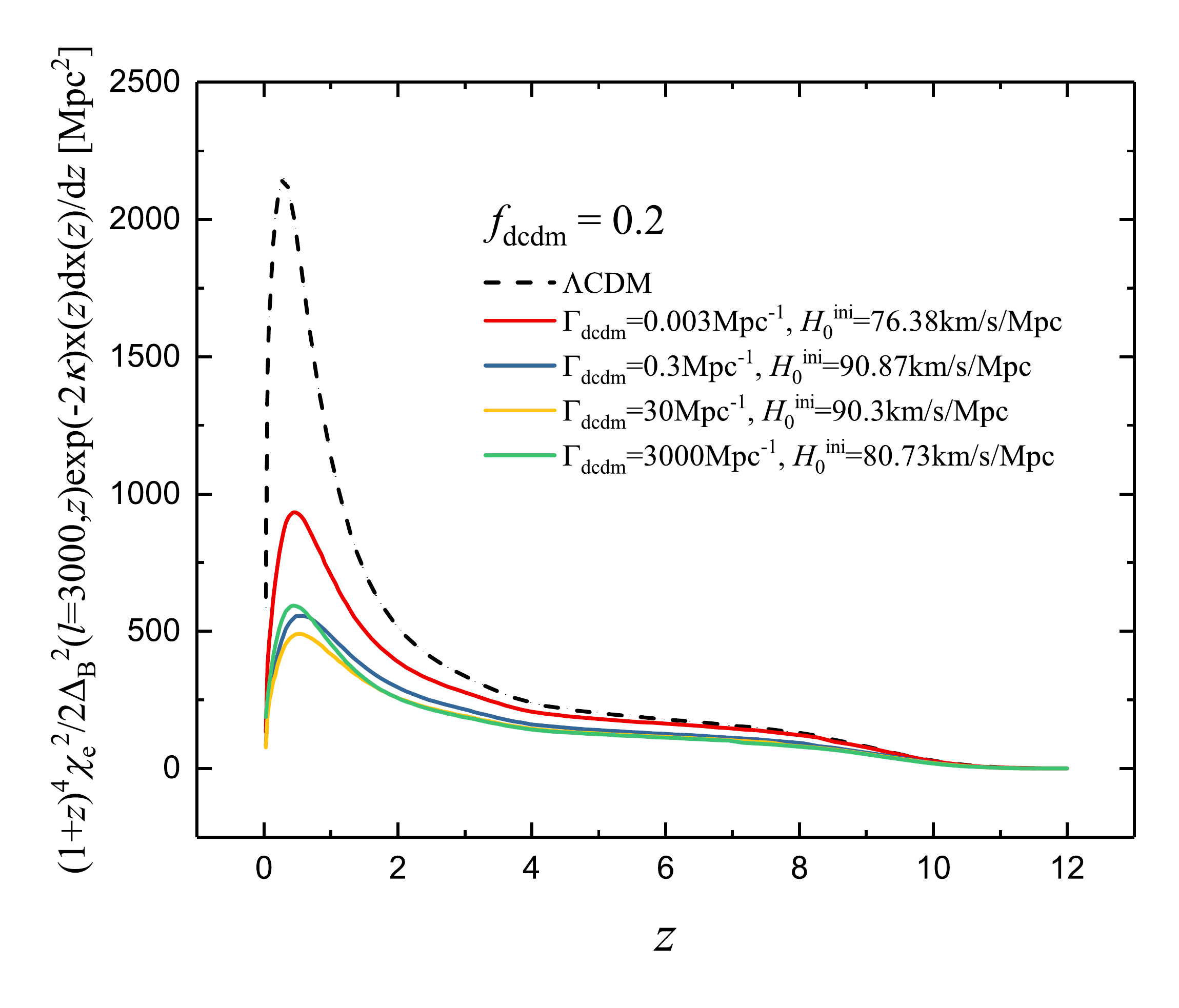}\label{fig:integrand}}\\
\subfloat[]{
\includegraphics[scale=0.32]{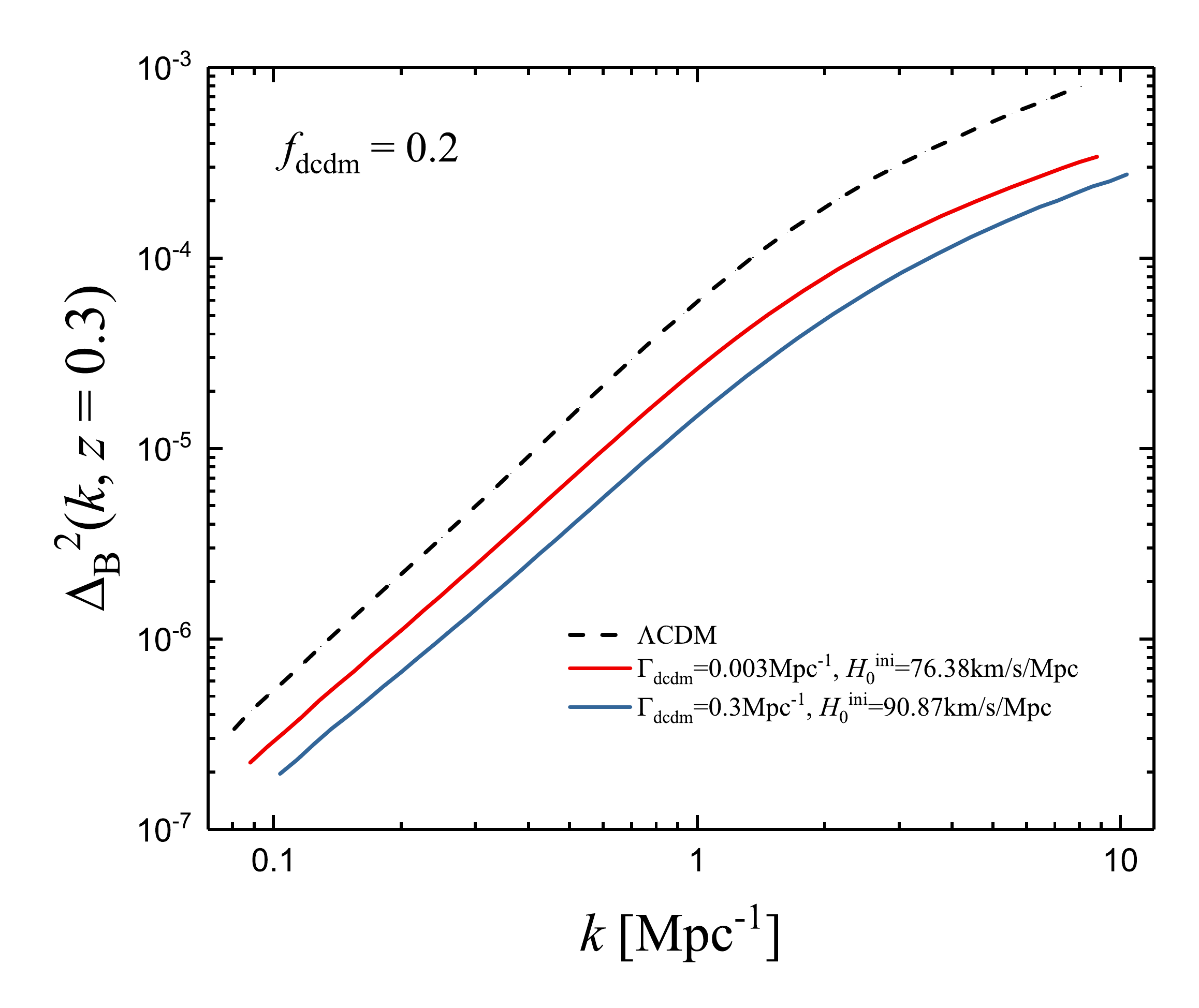}\label{fig:Delta_b2_1}}
\subfloat[]{
\includegraphics[scale=0.32]{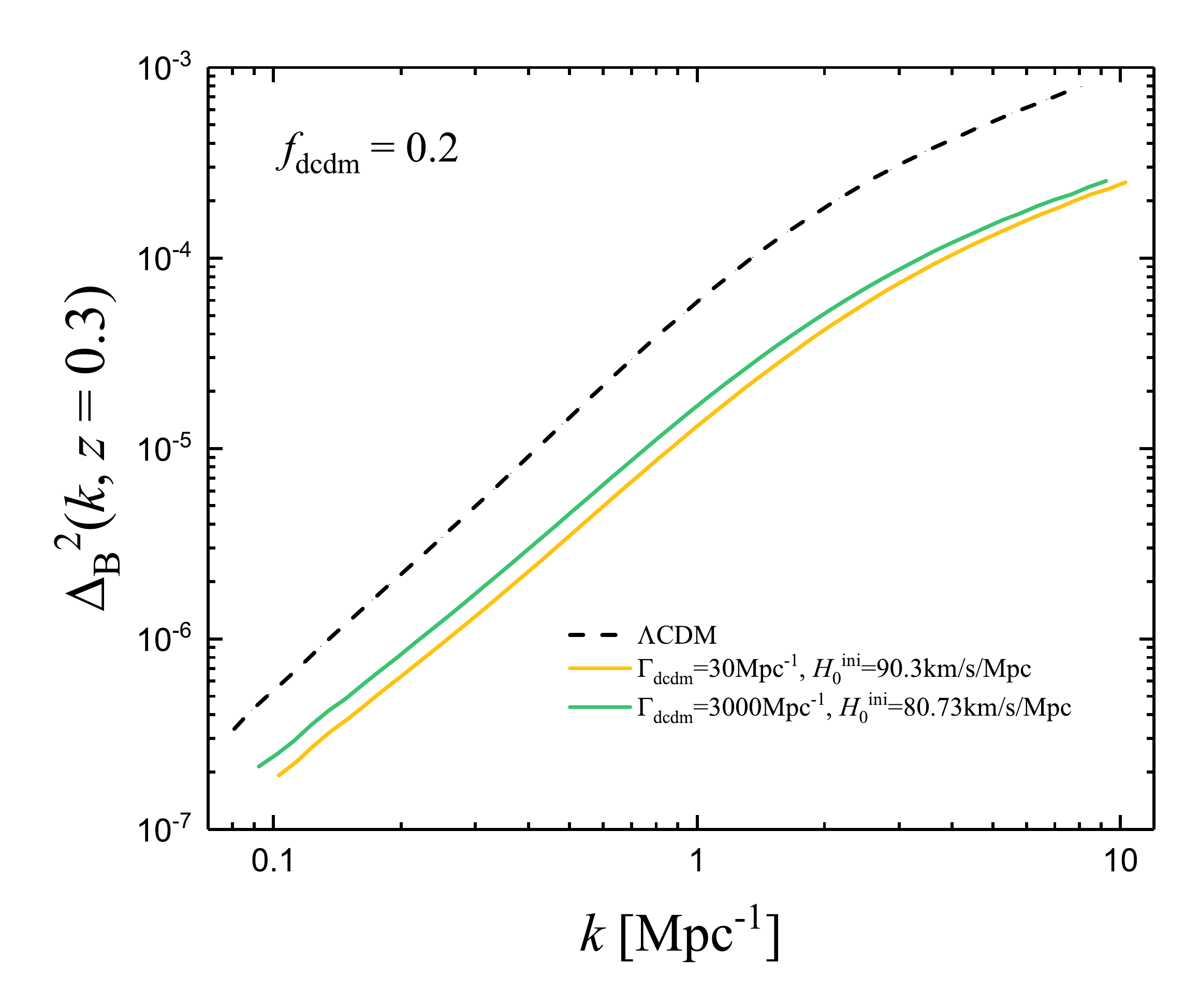}\label{fig:Delta_b2_2}}
\caption{(a) The integrand in Eq.~\ref{eq:cl_kSZ} as the function of the redshift $z$. As seen, the contributions from the low redshifts of $z\lesssim2$ will dominate the kSZ power spectrum $C_\ell^{kSZ}$, and thus the resulting $C_\ell^{kSZ}$ is almost insensitive to reionization history. The dimensionless power spectra of $\Delta_{B}^{2}$ in the presence of long-lived (bottom-left) and short-lived (bottom-right) DCDM have relatively smaller amplitudes than the $\Lambda$CDM-predicted one.}
\end{figure}

Fig.~\ref{fig:ClkSZ} shows the predicted angular power spectrum of the kSZ anisotropies in the presence of the DCDM, in comparison with the measured data from the South Pole Telescope (SPT)~\cite{George:2014oba}. The measured amplitude of the kSZ power spectrum by SPT is $2.9\pm1.3\mu{\rm K}^2$ at $\ell=3000$ and the corresponding $2\sigma$ upper limit is about $4.9\mu{\rm K}^2$. As seen, the predicted amplitude of kSZ from the standard $\Lambda$CDM is consistent within measured value at the $2\sigma$ level, whereas the current measurement in fact favour the DCDM model as the kSZ signal will be suppressed considerably by the presence of the DCDM which leads to the kSZ amplitudes consistent with data within $1\sigma$. Furthermore, the DCDM-induced influences in kSZ for the long-lived (in Fig.~\ref{fig:ClkSZ_1}) and short-lived (in Fig.~\ref{fig:ClkSZ_2}) models are essentially compatible with those in the bulk flow of the baryon shown in Fig.~\ref{fig:bf}.  

In Fig.~\ref{fig:integrand}, we illustrate the integrand in Eq.~\ref{eq:cl_kSZ} as the function of $z$ at $\ell = 3000$. One can find that the dominated contributions to the kSZ power spectrum only come from low redshifts of $z\lesssim2$, which means the kSZ signal is not strongly dependent on the reionization history.
Again, $100\theta_{\rm MC}$ is fixed in calculating the kSZ signals for different decay rates. With such ``fixing'', changing the decay rate will lead to a different dark energy density and Hubble expansion rate, and, consequently, leave an impact on $x(z)$ and $dx(z)/dz$ in Eq.~\ref{eq:cl_kSZ}. Furthermore, as shown in Fig.~\ref{fig:Delta_b2_1}, the power spectrum $\Delta_{B}^2(k,z)$ will decrease monotonically as increasing $\Gammadcdm$ in the long-lived case. From Fig.~\ref{fig:Delta_b2_2}, we also find that a strong suppression in $\Delta_{B}^2(k,z)$ appears whereas the strength of such suppression is insensitive to $\Gammadcdm$ in the short-lived case.   

With the above discussions about the DCDM-induced suppression in the kSZ power spectrum, the recent kSZ measurement by SPT would provide a further constraint on the parameter space of the DCDM model and MCMC-based constraints by using current observations are present in Sec.~\ref{sec:results}.

\section{Constraints on DM decay rate and fraction}\label{sec:results}
In this section, by running Monte Carlo Markov chains (MCMC) with using the public code CosmoMC~\cite{Lewis:2002ah,Lewis:2013hha}, we perform a joint analysis of the cosmological data to place tight constraints on the DCDM model. We compare the constraining power of various observables including the CMB, BAO, RSD and kSZ measurements as follows.

\subsection{Constraints from the CMB power spectra, BAO and $f\sigma_8$ data}
To understand how the $f\sigma_8$ measurements improve the constrains on the DCDM model, we start with using Planck 2015 temperature and polarization data over the full range of multipoles (referring to them as ``Planck2015'') \footnote{The CMB data used in the analysis include the Planck low-$\ell$ ($\ell<30$) TT, TE, EE spectra, combined with the high-$\ell$ data in the range of $\ell=30-2508$ for TT and $\ell=30-1996$ for TE and EE.}. We then include the $f\sigma_8$ data listed in Tab.~\ref{table:RSD} and refer to these data as ``RSD'' \footnote{Compared with Ref.~\cite{Chudaykin:2017ptd}, in this study we further take into account the decay before the recombination and perform an analysis with more RSD data involved.}. After that, we add BAO data from Six-degree-Field Galaxy Survey (6DFGS) at $z=0.106$~\cite{Beutler:2011hx}, SDSS Main Galaxy Sample (MGS) at $z=0.15$~\cite{Ross:2014qpa}, the BOSS-LOWZ at effective redshift $z_{eff}=0.32$ and the CMASS-DR11 at effective redshift $z_{eff}=0.57$~\cite{Anderson:2013zyy}, for placing further constraints on the DCDM model. In our analysis, we assume flat priors on the six cosmological parameters and two DCDM-related ones: $\{\omega_{b}^{ini}, \omega^{ini}_{c}, 100\theta_{\rm MC},\tau_{\rm reio}, n_s, A_s, \fdcdm, \Gammadcdm\}$. Note that, we choose a flat prior on $\Gammadcdm$ as $0\leq\Gammadcdm\leq5$ Mpc$^{-1}$ for the long-lived case and $0.5\leq\log_{10}(\Gammadcdm/$Mpc$^{-1})\leq3.5$ for the short-lived one. We also keep the total neutrino mass, the relativistic number of degrees of freedom and the spectrum lensing normalization as well as the Helium fraction fixed to $\Sigma m_\nu=0.06$eV, $N_{\rm eff}=3.046$, $A_{L}=1$ and $Y_{\rm He}=0.24$, respectively. We set a Gelman and Rubin criterion of $R-1=0.03$ to ensure the chains to be converged.

\begin{table*}
	\centering
	\caption{Best fit values and $68\%$ confidence levels for the cosmological parameters in long-lived DCDM cases. Note that, the current datasets cannot provide any constraints on $\Gammadcdm$.}
	\label{table:bf_s}
	\begin{tabular}{ccccccccc}
		\hline
		& \multicolumn{2}{c}{Planck2015+BAO} & \multicolumn{2}{c}{Planck2015+BAO+RSD} \\
		\cline{2-5}
		Parameter & Best fit  & 68\% limits & Best fit  & 68\% limits \\
		\hline
		$\omega_b^{ini}$ & 0.02233 & $0.02218^{+0.000157}_{-0.000156}$ & 0.02223 & $0.02218^{+0.00016}_{-0.00016}$\\
		$\omega_c^{ini}$ & 0.1189 & $0.12^{+0.00118}_{-0.00119}$ & 0.1198 & $0.1196^{+0.0012}_{-0.00121}$\\
		$100\theta_{MC}$ & 1.041 & $1.041^{+0.000308}_{-0.000305}$ & 1.041 & $1.041^{+0.000306}_{-0.000305}$\\
		$\tau$ & 0.09105 & $0.08297^{+0.0162}_{-0.016}$ & 0.07504 & $0.06615^{+0.0152}_{-0.0154}$\\
		$n_s$ & 0.9655 & $0.9647^{+0.00412}_{-0.00416}$ & 0.9635 & $0.9647^{+0.00412}_{-0.00408}$\\
		${\rm{ln}}(10^{10}A_s)$ & 3.114 & $3.101^{+0.0321}_{-0.0317}$ & 3.084 & $3.066^{+0.0297}_{-0.0301}$\\
        $\Gammadcdm/$Mpc$^{-1}$ & $-$ & $-$ & $-$ & $-$\\
        $\fdcdm$ & 0.001585 & $<0.008435$ & 0.008991 & $<0.009402$\\
        \hline
        $\Omega_\Lambda$ & 0.6917 & $0.6904^{+0.00681}_{-0.00678}$ & 0.6933 & $0.693^{+0.00665}_{-0.0066}$\\
        $\Omega_m$ & 0.3083 & $0.3096^{+0.00678}_{-0.00681}$ & 0.3067 & $0.307^{+0.0066}_{-0.00665}$\\
        $z_{\rm re}$ & 11.09 & $10.39^{+1.55}_{-1.3}$ & 9.7 & $8.817^{+1.57}_{-1.33}$\\
        $H_0$ & 67.83 & $67.93^{+0.531}_{-0.633}$ & 68.21 & $68.13^{+0.54}_{-0.64}$\\
        $\sigma_8$ & 0.836 & $0.8339^{+0.0132}_{-0.0131}$ & 0.8251 & $0.8181^{+0.0118}_{-0.0118}$\\
        ${\rm{Age}}/{\rm{Gyr}}$ & 13.79 & $13.78^{+0.0338}_{-0.0247}$ & 13.77 & $13.77^{+0.0347}_{-0.0264}$\\
		\hline
	\end{tabular}
\end{table*}

\begin{figure}[htbp!]
	\centering
	\subfloat[1D marginalized posterior distributions using Planck2015 data]{
		\includegraphics[scale=0.45]{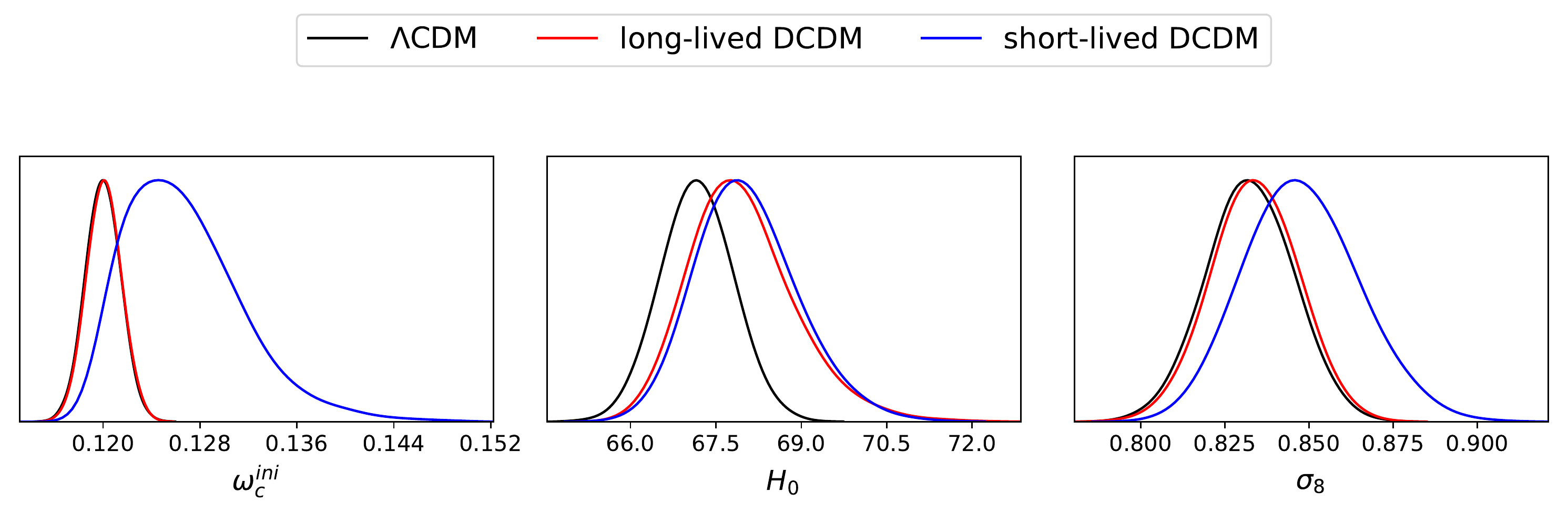}\label{fig:1D_plc}}\\
	\subfloat[1D marginalized posterior distributions using RSD data]{
		\includegraphics[scale=0.45]{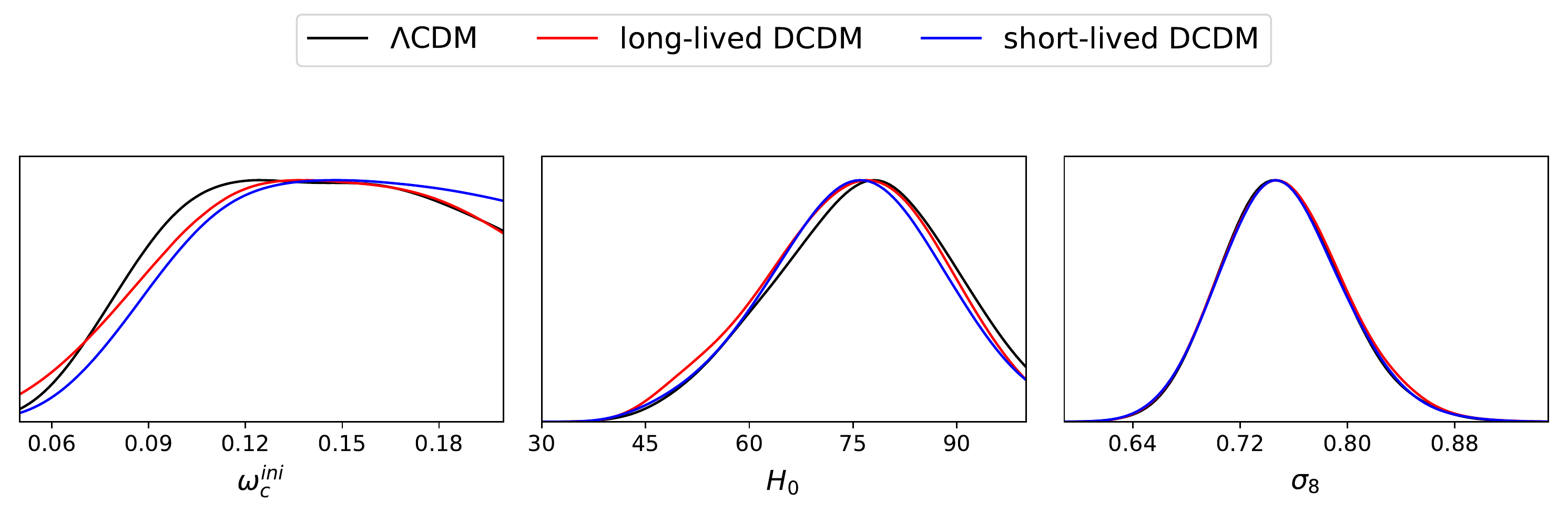}\label{fig:1D_RSD}}
	\caption{\label{fig:1D_distri} The comparison of 1D marginalized posterior distributions for $\omega_c^{ini}, H_0, \sigma_8$, with using either ``Planck2015'' or RSD data, respectively. In the $\Lambda$CDM model, there exist about $3\sigma$ tensions on $H_0$ and $\sigma_8$, but the DCDM model here cannot reduce such tensions significantly.}
\end{figure}

\begin{figure}
	\centering
	\includegraphics[scale=0.28]{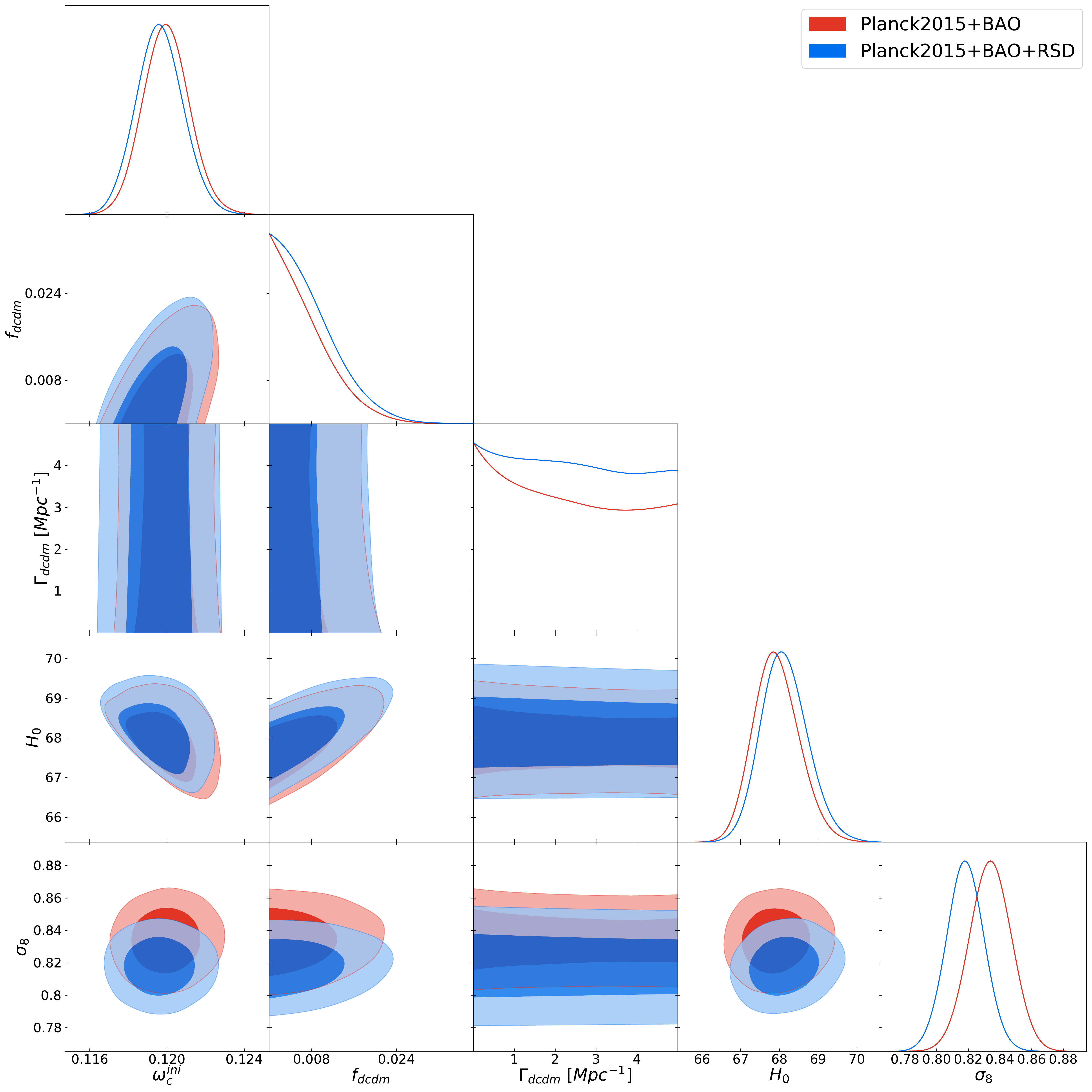}
	\caption{\label{fig:tri_s} 1D marginalized posterior distributions and 2D contours at $68\%, 95\%$ confidence levels for the parameters $\omega^{ini}_{\rm c}, \fdcdm, \Gammadcdm, H_0$ and $\sigma_8$, using the combined datasets of either ``CMB+BAO'' or ``CMB+BAO+RSD'' measurements in the long-lived DCDM model, respectively.}
\end{figure}

\begin{table*}
	\centering
	\caption{Best fit values and $68\%$ confidence levels for the cosmological parameters in short-lived DCDM cases.}
	\label{table:bf_b}
	\begin{tabular}{ccccccccc}
		\hline
		& \multicolumn{2}{c}{Planck2015+BAO} & \multicolumn{2}{c}{Planck2015+BAO+RSD} \\
		\cline{2-5}
		Parameter & Best fit  & 68\% limits & Best fit  & 68\% limits \\
		\hline
		$\omega_b^{ini}$ & 0.02224 & $0.02227^{+0.000165}_{-0.000164}$ & 0.02216 & $0.02226^{+0.000162}_{-0.000161}$\\
		$\omega_c^{ini}$ & 0.1299 & $0.1265^{+0.00273}_{-0.00644}$ & 0.1213 & $0.1227^{+0.00141}_{-0.00416}$\\
		$100\theta_{MC}$ & 1.041 & $1.041^{+0.000382}_{-0.000295}$ & 1.041 & $1.041^{+0.000338}_{-0.000301}$\\
		$\tau$ & 0.08943 & $0.085^{+0.0169}_{-0.0169}$ & 0.05536 & $0.06519^{+0.0156}_{-0.0155}$\\
		$n_s$ & 0.9787 & $0.9724^{+0.00548}_{-0.00768}$ & 0.9696 & $0.9685^{+0.00436}_{-0.00542}$\\
		${\rm{ln}}(10^{10}A_s)$ & 3.128 & $3.111^{+0.0338}_{-0.0339}$ & 3.052 & $3.066^{+0.0301}_{-0.0301}$\\
		${\rm{log_{10}}}(\Gammadcdm/$Mpc$^{-1})$ & 2.365 & $>2.245$ & 2.043 & $>2.087$\\
		$\fdcdm$ & 0.05596 & $0.04353^{+0.00906}_{-0.0426}$ & 0.01446 & $<0.02732$\\
		\hline
		$\Omega_\Lambda$ & 0.6741 & $0.6804^{+0.00916}_{-0.00736}$ & 0.6874 & $0.6865^{+0.00805}_{-0.0068}$\\
		$\Omega_m$ & 0.3259 & $0.3196^{+0.00736}_{-0.00916}$ & 0.3126 & $0.3135^{+0.0068}_{-0.00805}$\\
		$z_{\rm re}$ & 11.08 & $10.58^{+1.62}_{-1.33}$ & 7.832 & $8.717^{+1.59}_{-1.37}$\\
		$H_0$ & 68.46 & $68.37^{+0.605}_{-0.889}$ & 67.91 & $68.15^{+0.521}_{-0.694}$\\
		$\sigma_8$ & 0.8657 & $0.8486^{+0.0163}_{-0.0187}$ & 0.8167 & $0.8223^{+0.0128}_{-0.0129}$\\
		${\rm{Age}}/{\rm{Gyr}}$ & 13.68 & $13.72^{+0.0714}_{-0.0428}$ & 13.78 & $13.76^{+0.0472}_{-0.0272}$\\
		\hline
	\end{tabular}
\end{table*}

\begin{figure}
	\centering
	\includegraphics[scale=0.28]{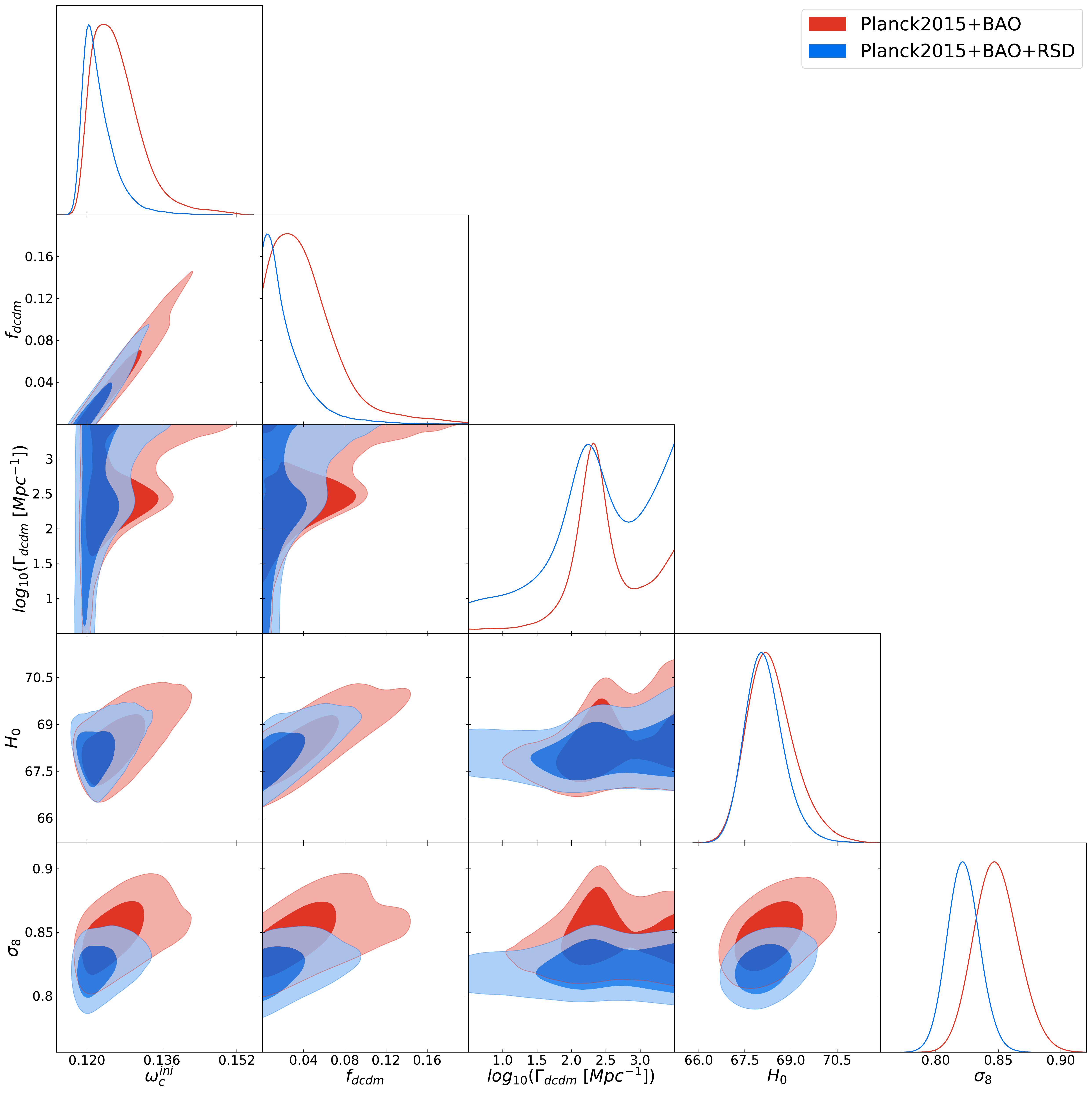}
	\caption{\label{fig:tri_b} Same as in Fig.~\ref{fig:tri_s}, but for the short-lived DCDM model.}
\end{figure}

In Fig.~\ref{fig:1D_distri}, we first consider the models of $\Lambda$CDM and DCDM constrained by ``Planck2015'' and RSD data, respectively. We see that, the inferred values of $H_0$ and $\sigma_8$ from CMB data are still in about $3\sigma$ tensions with those from RSD data and the DCDM model can only slightly alleviate such tensions, which confirms the results found in Ref.~\cite{Poulin:2016nat}. Then in Table.~\ref{table:bf_s}, we present the results for the long-lived DCDM model from the combined datasets ``Planck15+BAO'' and ``Planck15+BAO+RSD'', and the inferred 1D distributions and 2D contours are shown in Fig.~\ref{fig:tri_s}.

Using the ``Planck15+BAO'' dataset, we find a $68\%$ CL upper limits on the fraction is $\fdcdm\lesssim0.84\%$ but very weak constraint on $\Gammadcdm$ due to the tiny changes from the decay in the CMB angular power spectra at $\ell \gtrsim 100$ (see Fig.~\ref{fig:CMB1} and Fig.~1 in Ref.~\cite{Poulin:2016nat}). When including RSD data in the analysis, one can not obtain stronger constraint on $\Gammadcdm$, and the bound on $\fdcdm$ now becomes weaker by about a factor of 1.1. These results indicate that the RSD data prefer a relatively larger decaying DM fraction, consistent with our findings in Sec.~\ref{sec:growth} where the predicted values of $f\sigma_8$ fit the data better than the $\Lambda$CDM predicted. Also, as expected, the current datasets can not effectively break down the degeneracy between $\Gammadcdm$ and $\fdcdm$ for the long-lived model (see Fig.~\ref{fig:CMB_f}).

The constraints in the short-lived DCDM model (most of the DCDM has decayed away before the recombination) are summarized in Tab. \ref{table:bf_b} and the corresponding 1D marginalized posterior distributions and 2D contours are shown in Fig.~\ref{fig:tri_b}. To speed up the exploration of the parameter space in $\Gammadcdm$, we scan over $\log_{10}(\Gammadcdm/$Mpc$^{-1})$ with a flat prior instead of $\Gammadcdm$. Due to the fact that the total DM density at the recombination is tightly fixed by the CMB and increasing $\fdcdm$ will lead to the total DM density significantly decreased at that time, therefore one can find a strong positive correlation between initial DM density $\omega_c^{ini}$ and DM fraction $\fdcdm$.  From the 1D distributions, we also find that, the RSD data provides strong constraints on $\fdcdm$ and $\omega_c^{ini}$ and the bound on $\Gammadcdm$, however, becomes slightly weaker than that by ``Planck2015+BAO''. As known from Fig.~\ref{fig:fs8_2}, the predicted $f\sigma_8$ from the short-lived case is noticeably lower than the $\Lambda$CDM one, and thus the current RSD data obtained from the surveys in late universe are much more in favor of the lower abundance of DCDM. Furthermore, Fig.~\ref{fig:fs8_2} also indicates that for a fixed $\fdcdm$ the larger $\Gammadcdm$ leads to a smaller suppression on $f\sigma_8$ and gives a better fit to the data. As a result, a lower abundance of DCDM with a relatively larger decay rate would be also compatible with the data ``Planck2015+BAO+RSD'', which gives rise to a relatively weaker bound on $\Gammadcdm$ with the inclusion of RSD data. Finally, the marginalized $1\sigma$ upper limit on $\fdcdm$ is improved by a factor of about 1.9 ($\lesssim5.26\%$ from ``Planck2015+BAO'' and $\lesssim2.73\%$ from ``Planck2015+BAO+RSD''), when the RSD data are included in the analysis.

\subsection{Adding kSZ data from SPT}
\begin{figure}[!htb]
\centering
\includegraphics[scale=0.21]{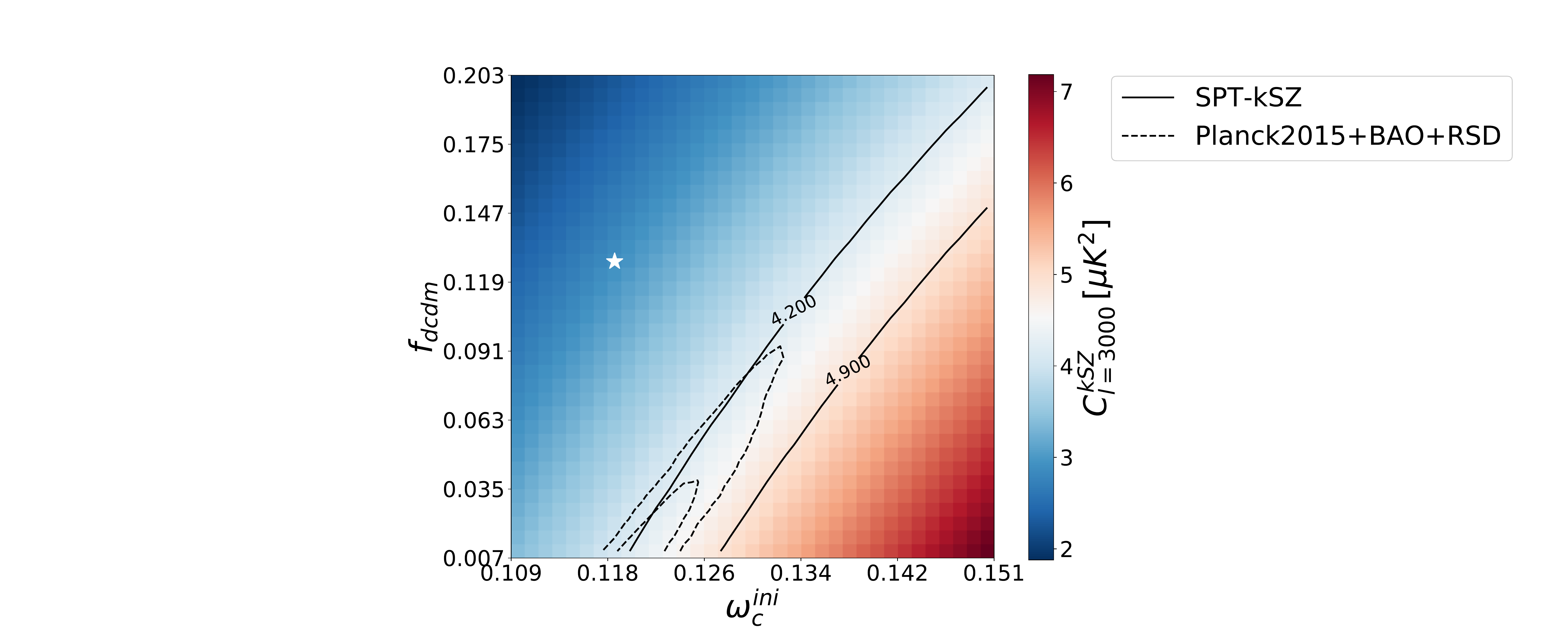}\\
\includegraphics[scale=0.21]{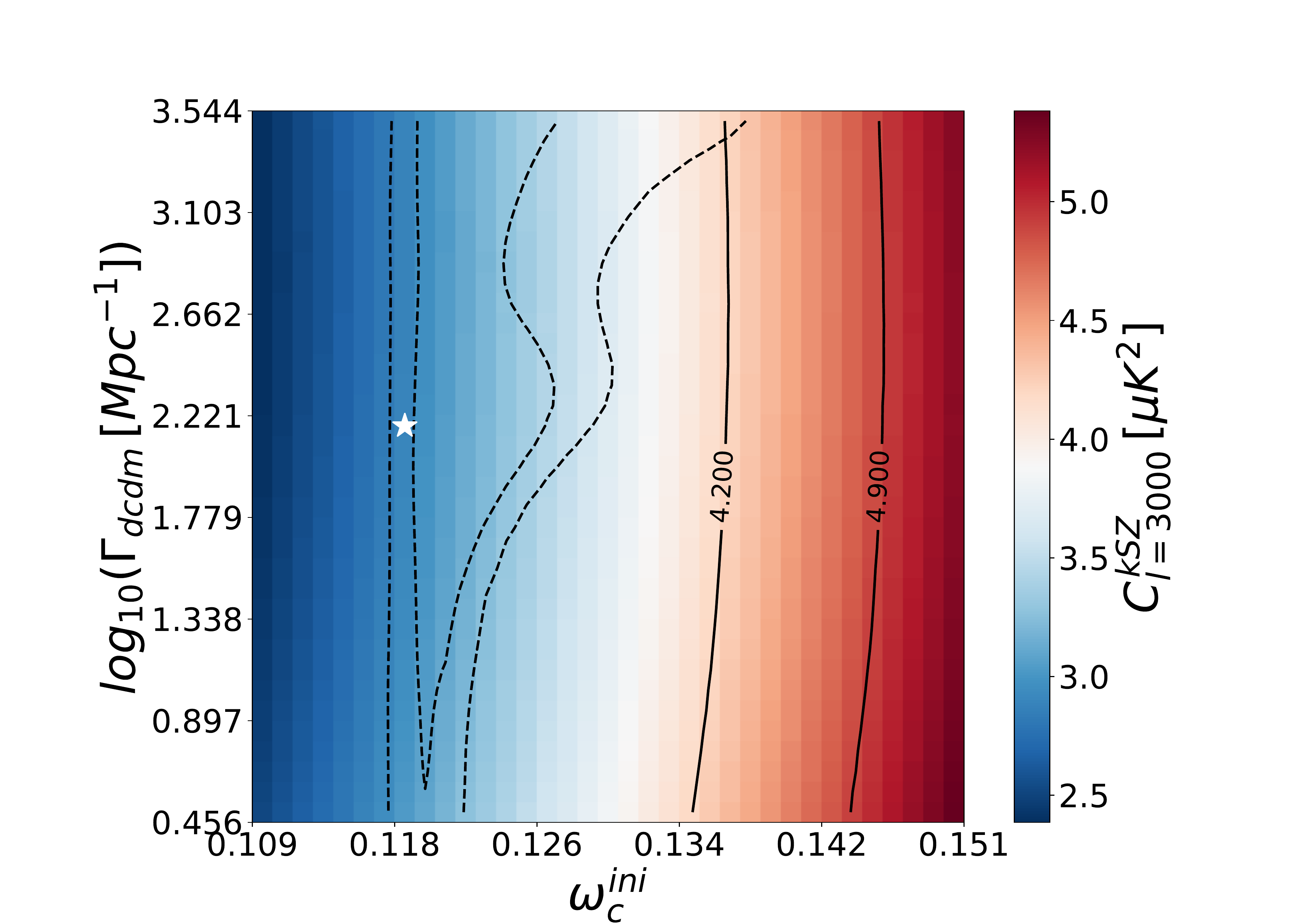}
\includegraphics[scale=0.21]{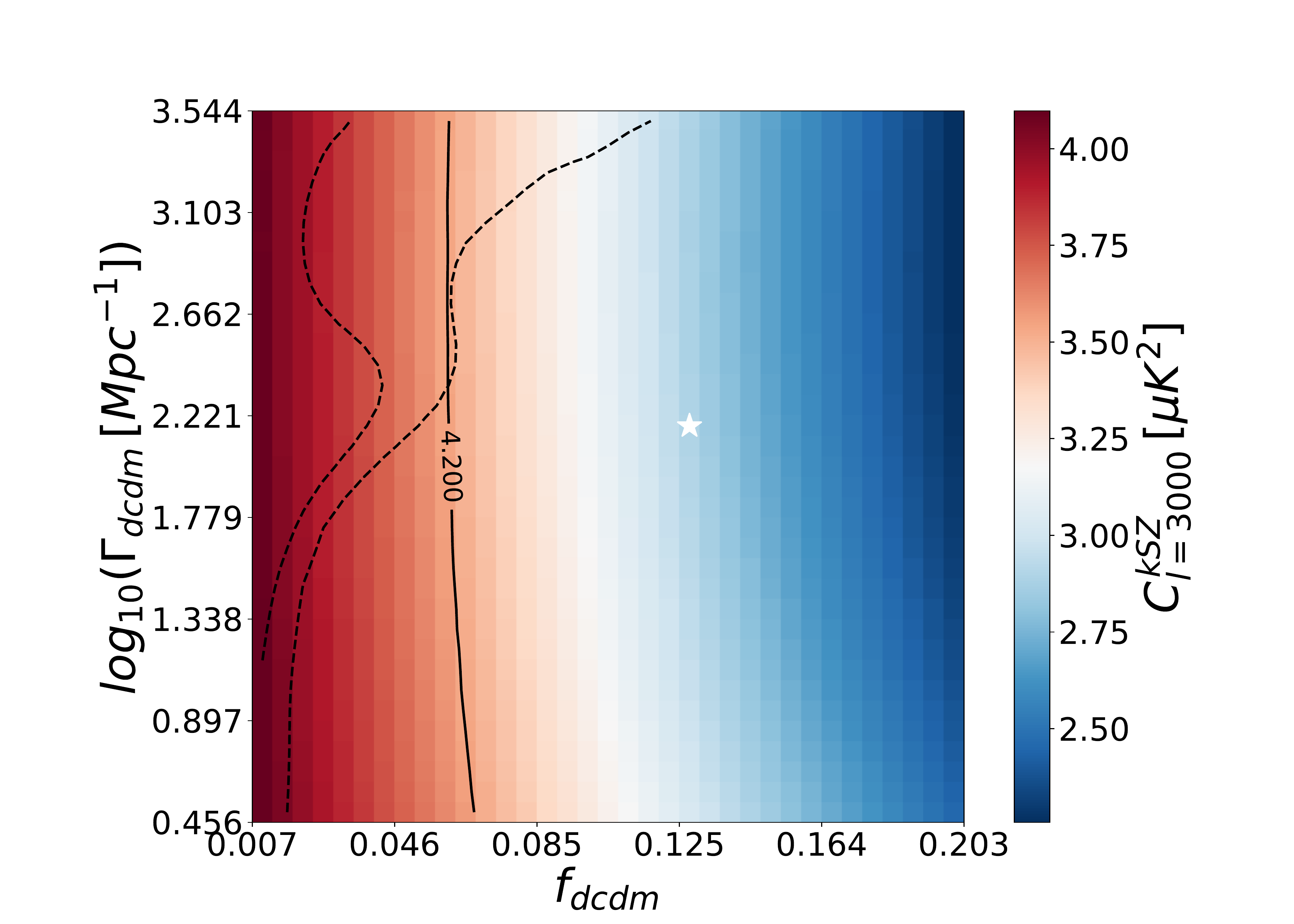}\\
\caption{\label{fig:kSZ_cont_b} Comparison of the constraints on the parameters $\omega^{ini}_{\rm c}, \fdcdm$ and  $\Gammadcdm$  for the short-lived DCDM model, from the SPT measurement of the kSZ signal at $\ell=3000$ and from the ``Planck2015+BAO+RSD'' dataset. The solid lines denote the SPT-measured $1\sigma$ upper limit of $4.2\mu{\rm K}^2$ and $2\sigma$ upper limit of $4.9\mu{\rm K}^2$, and the dashed lines are the ``Planck2015+BAO+RSD''-induced contours from Fig.~\ref{fig:tri_b}. The white star corresponds to the best-fitted $C_{l=3000}^{kSZ}=2.9\mu$K$^2$ by SPT.} 
\end{figure}

\begin{figure}[!htb]
\centering
\includegraphics[scale=0.21]{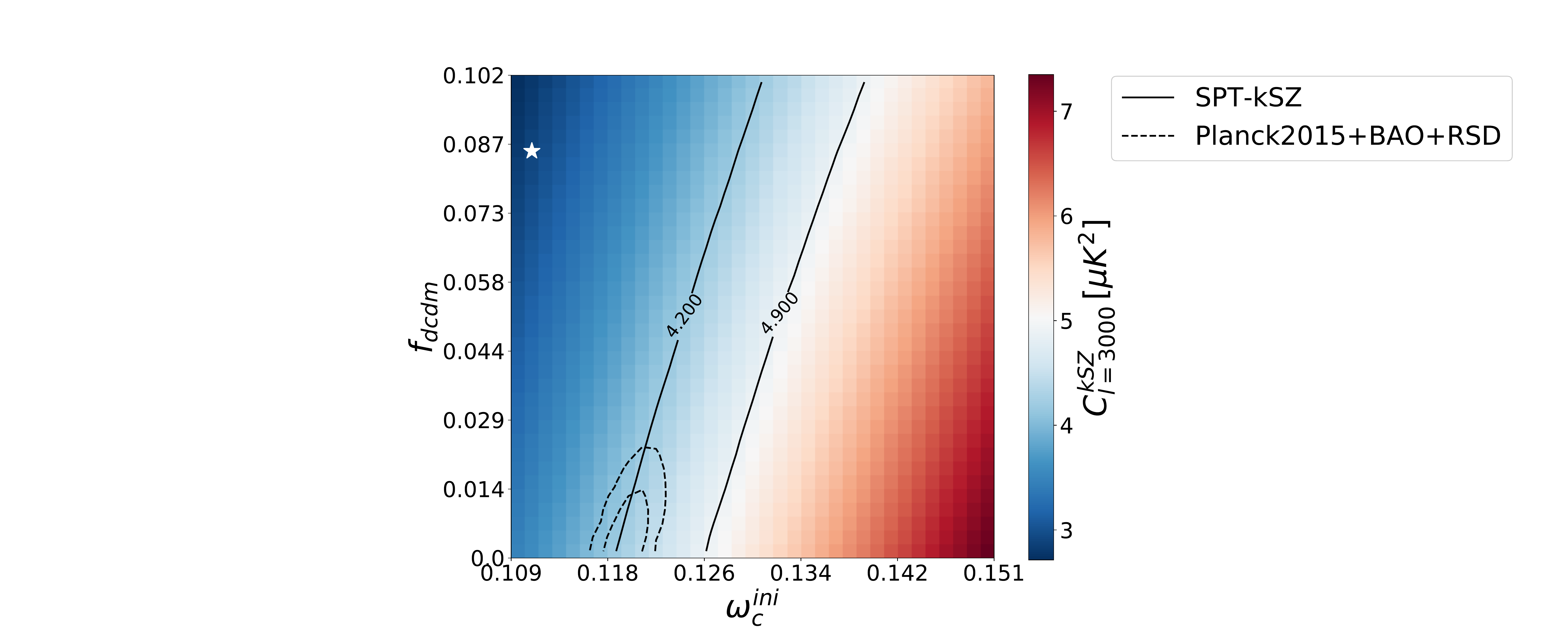}\\
\includegraphics[scale=0.21]{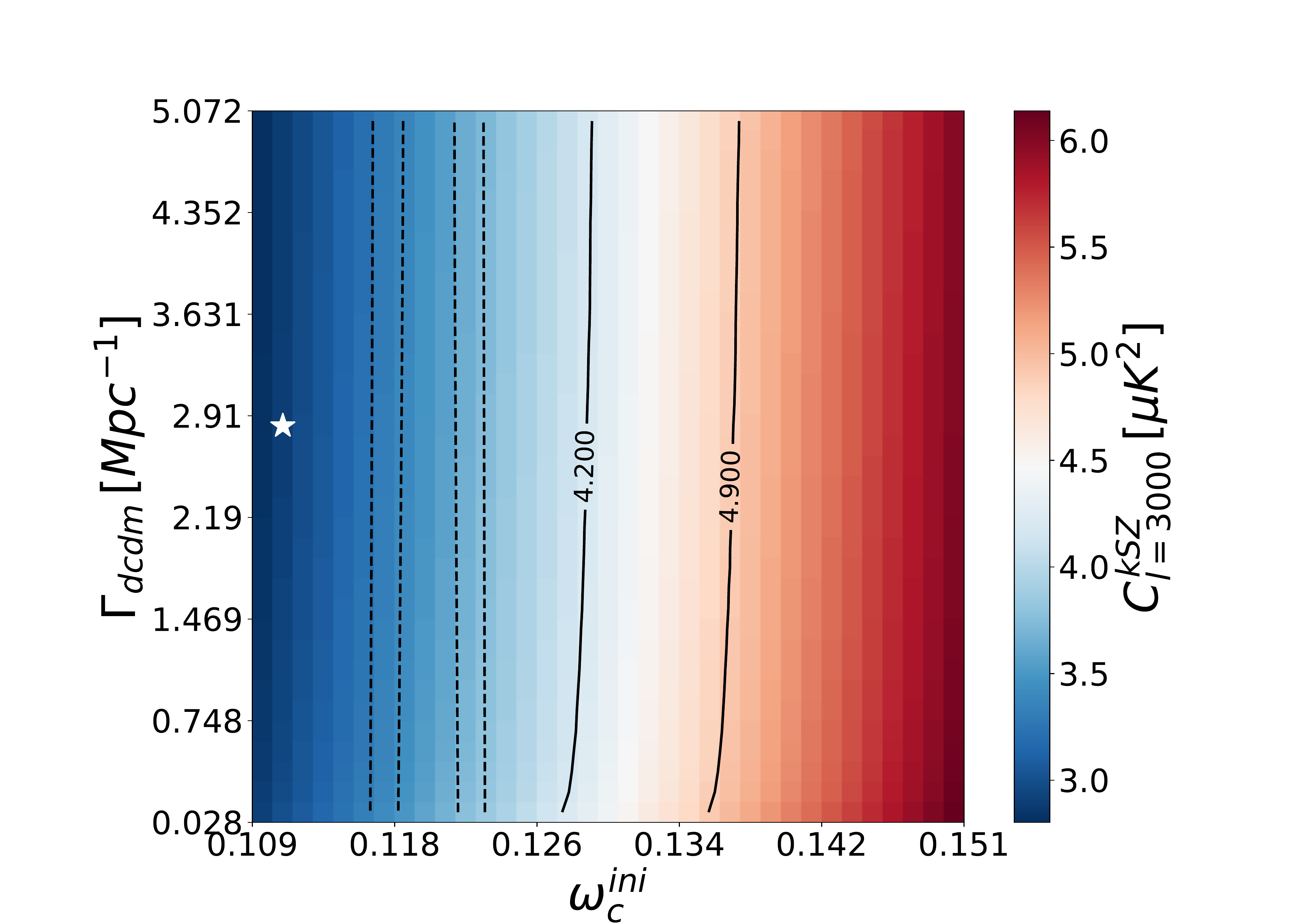}
\includegraphics[scale=0.21]{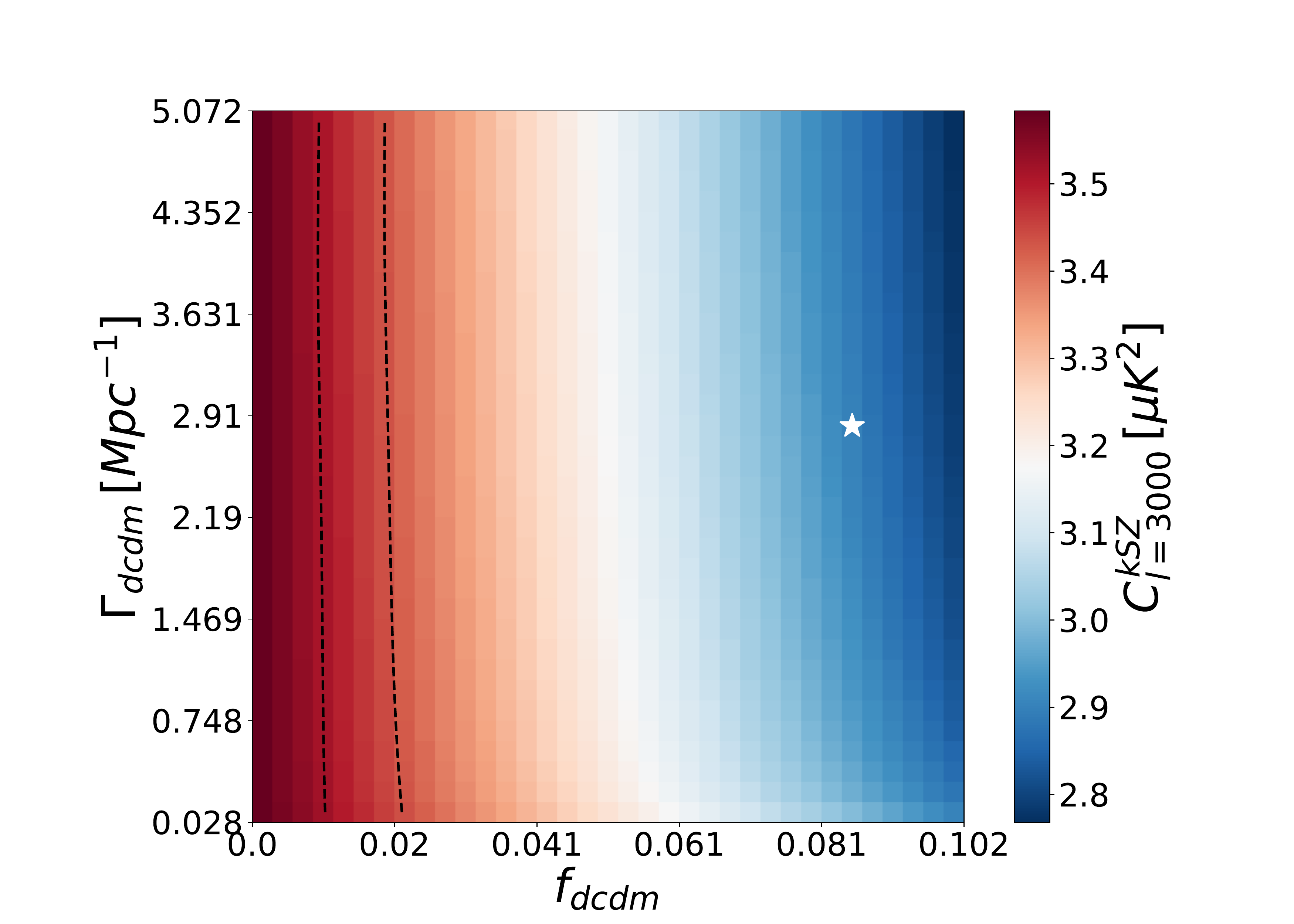}\\
\caption{\label{fig:kSZ_cont_s} Same as Fig.~\ref{fig:kSZ_cont_b}, but for the long-lived DCDM model.} 
\end{figure}

The influences of DM decays on the kSZ signal has been investigated in Sec.~\ref{sec:kSZ}, and thus let us compute the constraining power of the SPT-kSZ data. Since the SPT provides a measurement of the kSZ signal at $\ell =3000$ only,  it is not able to provide significant improvements in the constraints on the DCDM model. Here, we therefore only investigate the constraint from such kSZ data by fixing cosmological parameters with the ``Planck 2015'' best-fitted values but varying $\omega_{c}^{ini}$, $\fdcdm$ and $\Gammadcdm$ that allows us to construct a 3D parameter space. In Fig.~\ref{fig:kSZ_cont_b} and~\ref{fig:kSZ_cont_s}, we show the results for the long- and short-lived DCDM models, respectively. Comparing with the constraining power of the ``Planck2015+BAO+RSD'' dataset, the current kSZ data prefers a non-zero fraction of DCDM, which is consistent with the results in Sec.~\ref{sec:kSZ} but somewhat opposite to the preference from ``CMB+BAO+RSD'' measurements. We find that the best fitted value of the SPT-kSZ measurement (the ``white star'' in Fig.~\ref{fig:kSZ_cont_b} \&~\ref{fig:kSZ_cont_s}) deviates from the ``Planck2015+BAO+RSD''-inferred one at 2$\sigma$ level. This disagreement might imply a possible tension between the current kSZ and CMB data, and future precise kSZ measurements would provide independent evidence for the existence of DCDM.

\section{Conclusions}
\label{sec:conclu}
In this work, we have investigated cosmological bounds on a fractional decaying DM that is allowed to decay into invisible relativistic particles (``dark radiation''). The DCDM model is characterized by two parameters, the fraction $\fdcdm$ of the DCDM and the decay width $\Gammadcdm$, and we estimate the constraints on such two parameters by purely gravitational effects from the measurements of the CMB temperature anisotropies, the redshift distortion and the kinetic Sunyaev-Zel'dovich effect. Based on a full calculation of the evolution equations for the background and the perturbations by modifying the full Boltzmann hierarchy, we have quantitatively determined the impacts of the decaying DM on cosmological observables and placed a cosmological constrains on $\fdcdm$ and $\Gammadcdm$ from a joint analysis of the cosmological data by running Monte Carlo Markov chains.

With respect to the previous literature, the main features of this paper are as follows. 1) More cosmological observables have been taken into account in the analysis, including not only the CMB temperature anisotropies, but also the structure growth rate, the bulk flow of the baryon and the kSZ effect. 2) We have evaluated in detail the impacts of the DM decay on the Sachs-Wolfe (SW) effect, the early/late Integrated SW effects, as well as on the evolution of the Weyl potential $|\Phi|$, which explicitly provides physical explanations for the decay-induced signatures in the CMB TT power spectrum, the growth rate $f\sigma_8$, the bulk flow of baryon $\langle v_{\mathrm{b}}^2 \rangle^{1/2}$, and the kSZ power spectrum. We find that the DM decay leads to a visible suppression in $|\Phi|$ and consequently changes the above-mentioned cosmological observables accordingly. 3) When adding growth rate measurements of the RSD data, the $68\%$ CL upper limit tightens considerably from $\fdcdm\lesssim5.26\%$ to $\fdcdm\lesssim2.73\%$ for the short-lived DCDM model, whereas the inclusion of RSD data can not place a tighter constraint on $\fdcdm$ than that from the combined dataset of ``CMB+BAO'' for the long-lived one. our results also show that the DCDM model does not significantly alleviate the $H_0$ and $\sigma_8$ tensions, which remain at 3$\sigma$ level. 4) For the first time, we also investigated the constraining power of the kSZ effect on the DCDM model in detail. The current SPT-kSZ prefers the presence of the decaying DM that leads to a suppression in the kSZ power spectrum. The derived contours from the kSZ data alone, which are different from the  ``CMB+BAO+RSD'' inferred ones, indicate that the kSZ signal is sensitive to the DCDM model and would provide an independent constraint on the parameter space. We thus expect that future precise kSZ surveys will constrain the properties of DCDM tightly.

Definitely, if the cold DM decays into electromagnetic particles, many other cosmological observables will provide additional information for detecting/constraining the decaying DM, such as the reionization history (e.g.,~\cite{Oldengott:2016yjc,Rudakovskiy:2016ngi,Rudakovskyi:2018jfc}), and, the existence of DCDM can be strictly tested by ongoing and future reionization observations~\cite{Mellema:2012ht,Pober:2013jna,Patil:2014dpa,Ali:2015uua,Bowman:2012ef,Monsalve:2016xbk,Aghanim:1996ib,Gruzinov:1998un}. Moreover,  the weak lensing surveys~\cite{Hildebrandt:2016iqg,Laureijs:2011gra,Hikage:2018qbn,Abate:2012za} are expected to further improve the constraints considerably. Finally,  it is worthwhile to examine the impacts of DCDM models on the non-linear clustering regime carefully by N-body simulations in a future study.

\appendix
\section{Adjust the initial $H_{0}^{ini}$ and $\rho_\Lambda$ for keeping $\theta_{\rm MC}$ fixed}
\label{sec:app_A}
When investigating and discussing the new physics using cosmological observations, one has to specify which cosmological parameters are kept fixed, in order to clearly demonstrate predicted signatures from new physics.    

Specifically, the effects of DM decay rate $\Gammadcdm$ on the CMB can lead to a change in the angular diameter distance to decoupling and  a shift of the peaks and trough positions, compared with the standard $\Lambda$CDM predictions.  To cancel the trivial effects in the CMB and to discuss the effects associated to the decays conveniently, we keep a constant value of $\theta_{\rm MC}$ for the DCDM model before Sec.~\ref{sec:results} and, consequently, the initial Hubble parameter $H_{0}^{ini}$ and $\rho_\Lambda$ will be adjusted accordingly for a given $\Gammadcdm$ and $\fdcdm$. Note that, we however treat $\theta_{\rm MC}$ as a free parameter for the  MCMC analysis in Sec. \ref{sec:results}. 

Now, let us focus on the dependence of $H_{0}^{ini}$ and $\rho_\Lambda$ on $\Gammadcdm$ when fixing $\theta_{\rm MC}$. By definition, 

\begin{eqnarray}
\theta_{\rm MC} = d_s(z_*)/d_A(z_*)\,,  
\end{eqnarray}
where
\begin{eqnarray}
  &&d_s(z_*) = \int_{0}^{a_*} C_s\,\frac{da}{a\mathcal{H}(a)}\,,\\
  &&d_A(z_*) = \int_{a_*}^{1} \,\frac{da}{a\mathcal{H}(a)}\,.
\end{eqnarray}
Here,  $d_s(z_*)$ and $d_A(z_*)$ denote the sound horizon before the photon decoupling at $z_*$ and the angular diameter distance from last scattering surface to us, respectively. $C_s$ is the speed of sound in the photon-baryon fluid. As expected, the decay will change the Hubble expansion rate  $H(a) = \mathcal{H}(a)/a$ due to some amount of DM ($\propto a^{-3}$) decaying into invisible relativistic particles ($\propto a^{-4}$), and then alter the $\theta_{\rm MC}$. By assuming the present-day value of the baryon density is fixed, varying the dark energy density $\rho_\Lambda$ gives a solution for compensating for the change in $H(a)$.

For the long-lived model, as the typical time scale of the decay $\gg$ $1/H(z_*)$, the $d_s(z_*)$ is insensitive to the decay, but the $d_A(z_*)$ would increase significantly due to a large decrease in $H(z)$. We therefore increase $\rho_\Lambda$ to reduce the $d_A(z_*)$ so as to keep $\theta_{\rm MC}$ unchanged. Moreover, the larger $\rho_\Lambda$ will enhance the CMB low-$\ell$ power due to  the late integrated Sachs-Wolfe effect.

For the short-lived model where $\Gammadcdm \gtrsim 5$ Mpc$^{-1}$, the typical decay occurs before the recombination and only a small fraction of decaying DM are still alive at the late-time universe. In this scenario, the decay has impacts on both  $d_s(z_*)$ and $d_A(z_*)$, unlike the long-lived model. Increasing $\Gammadcdm$ will lead to both $d_s(z_*)$ and $d_A(z_*)$ increased as $H(z)$ decreased. However, $\theta_{\rm MC}$, the ratio of $d_s(z_*)$ and $d_A(z_*)$, does not monotonically change in $\Gammadcdm$, since the dependences of $d_s(z_*)$ and $d_A(z_*)$ on $\Gammadcdm$ are obviously different.  As a result, in some cases with considerably large $\Gammadcdm$, the changes in $\theta_{\rm MC}$ would become even smaller than those from a smaller $\Gammadcdm$ (shown in Fig.~\ref{fig:DCDM_BG}). 

In Tab.~\ref{tab:SAT}, we illustrate the derived values of $d_s(z_*)$, $d_A(z_*)$ and $100\theta_{\rm MC}$ by varying $\Gammadcdm$ and $\rho_{\Lambda}$, while keeping $\fdcdm=0.2$ fixed. For each $\Gammadcdm$, with adjusting $\rho_{\Lambda}$ to an appropriate value or without, we clearly demonstrate the effects of the decay in $\theta_{\rm MC}$. When fixing $\theta_{\rm MC}$ to the standard value $\theta^{\Lambda{\rm CDM}}_{\rm MC}$ (i.e., $\theta_{\rm MC}/\theta^{\Lambda{\rm CDM}}_{\rm MC}=1$), we qualitatively confirm the above claims. One can find that, the increase in $\rho_\Lambda$ for $\Gammadcdm = 3000$ Mpc$^{-1}$ (i.e., short-lived) is smaller than that for  $\Gammadcdm = 0.3$ or 30 Mpc$^{-1}$, and $\rho_\Lambda$ must be adjusted to the higher values (increased by 42.2$\%$ and by $220.5\%$ for $\Gammadcdm = 0.003$ and $0.3$ Mpc$^{-1}$ for the long-lived model, respectively.

\begin{table}[h!]
  \caption{Values of $d_s(z_*)$, $d_A(z_*)$ and $\theta_{\rm MC}$ by varying $\Gammadcdm$ and $\rho_{\Lambda}$} \centering \label{tab:SAT}
  \begin{tabular}{ccccc}
    \hline
    \hline
    $\Gammadcdm/$Mpc$^{-1}$& $\rho_{\Lambda}/\rho^{\Lambda \rm CDM}_{\Lambda}$  & $d_s(z_*)/$ Mpc$^{-1}$ & $d_A(z_*)/$Mpc$^{-1}$ & $\theta_{\rm MC}/\theta^{\Lambda{\rm CDM}}_{\rm MC}$  \\
    \hline
    0      &1           & 144.551     & 13889.245     & 1  \\
    0.003  &1.422       & 144.551     & 13888.883     & 1 \\
    0.003 &1           & 144.551     & 14274.113     & 0.973  \\
    0.3    &2.205       & 144.580     & 13891.647     & 1  \\
    0.3   &1           & 144.580     & 14847.220     & 0.935  \\
    30     &2.171       & 146.391     & 14065.403     & 1  \\
    30    &1           & 146.391     & 15002.749     & 0.937  \\
    3000   &1.643       & 150.297     & 14441.074     & 1  \\
    3000  &1           & 150.297     & 15019.707     & 0.961  \\
    \hline
    \hline
  \end{tabular}
\end{table}

\section{Behaviours of $|\Phi|$ on large $\Gammadcdm$}
\label{sec:app_B}
Here, we apply for a semi-analytical method to qualitatively understand the behaviours of $|\Phi|$ on large $\Gammadcdm$ mentioned in Sec.~\ref{sec:CMB}, where a larger $\Gammadcdm$ however leads to a smaller suppression in $\Phi$, completely different from what we would expect.

For simplicity (without loss of generality), suppose that $\Omega_m + \Omega_r=1$, Eqs.~\ref{eq:rhodcdm} and ~\ref{eq:rhodr}  then reads  
\begin{eqnarray}
\frac{d\rho_m}{da} &=& -\frac{3}{a}\rho_m-\frac{\Gamma_m}{\mathcal{H}}\rho_m\,,\label{eq:rho_m}\\
\frac{d\rho_r}{da} &=& -\frac{4}{a}\rho_r+\frac{\Gamma_m}{\mathcal{H}}\rho_m\,.\label{eq:rho_r}
\end{eqnarray}
The analytic solutions of these two equations are expressed by 
\begin{eqnarray}
\rho_m(a) &=& \frac{\rho_m^{ini}a_{ini}^3}{a^3}e^{-\Gamma_mt}\,,\\
\rho_r(a) &=& \frac{\rho_r^{ini}a_{ini}^4}{a^4} + \Gamma_m\frac{\rho_m^{ini}a_{ini}^3}{a^4}\int_{0}^{t}a'e^{-\Gamma_mt'}\,dt'\,,
\end{eqnarray}
where $t$ is the proper time and the (') denotes integration variables. The Weyl potential $|\Phi|$ is determined by the Poisson equation, 
\begin{eqnarray}
k^2|\Phi| = 4\pi Ga^2(\rho_m\Delta_m + \rho_r\Delta_r)\,,\label{eq:Poi}
\end{eqnarray}
in which $\Delta_m$ and $\Delta_r$ collect all the linear terms in perturbations (e.g., overdensity, peculiar velocity, etc.), and $\rho_m \propto a^{-3}$ and $\rho_r \propto a^{-4}$.

Recalling the adopted initial condition described in Sec.~\ref{sec:ICS_BG}, and defining $\Delta|\Phi| \equiv |\Phi|_{\rm dcdm}-|\Phi|_{\Lambda \rm CDM}$, the deviation of $|\Phi|$ from the $\Lambda \rm CDM$,  one can explicitly rewrite Eq.~\ref{eq:Poi} as 
\begin{eqnarray}\label{eq:Dphi}
  \begin{split}
    k^2\Delta|\Phi(a)| &= \frac{\rho_m^{ini}a_{ini}^3}{a^3}e^{-\Gamma_mt}\Delta_m + (\frac{\rho_r^{ini}a_{ini}^4}{a^4} + \Gamma_m\frac{\rho_m^{ini}a_{ini}^3}{a^4}\int_{0}^{t}a'e^{-\Gamma_mt'}\,dt')\Delta_r\\
    &\ \ \ \ -\frac{\rho_m^{ini}a_{ini}^3}{a^3}\Delta_m - \frac{\rho_r^{ini}a_{ini}^4}{a^4}\Delta_r\\
    &= \frac{\rho_m^{ini}a_{ini}^3}{a^3}[(e^{-\Gamma_mt}-1)\Delta_m + (\frac{\Gamma_m}{a}\int_{0}^{t}a'e^{-\Gamma_mt'}\,dt')\Delta_r]\\
    &= \frac{\rho_m^{ini}a_{ini}^3}{a^3}[(e^{-\Gamma_mt}-1)\Delta_m - \frac{1}{a}(a'e^{-\Gamma_mt'}\bigg|_{0}^{t} - \int_{0}^{t}e^{-\Gamma_mt'}\,da')\Delta_r]\\
    &= \frac{\rho_m^{ini}a_{ini}^3}{a^3}[(e^{-\Gamma_mt}-1)\Delta_m + (\frac{1}{a}\int_{0}^{a}e^{-\Gamma_mt'}\,da' - e^{-\Gamma_mt})\Delta_r]\,.
  \end{split}
\end{eqnarray}
 Here, we have assumed that, before the matter-dominated epoch there is no difference in $\Delta_m$ or $\Delta_r$ between the $\Lambda$CDM and the DCDM model. During the radiation dominated epoch (RD),  $t \sim 10^{-3}$ Mpc when $a \sim 10^{-4}$ and  $t \sim 10^{-9}$ Mpc when $a \sim 10^{-7}$, so that  $\frac{1}{a}\int_{0}^{a}e^{-\Gamma_mt'}\,da' \simeq 1$. With this approximation, one can further derive the following equation, 
\begin{eqnarray}\label{eq:Dphi_apprx}
k^2\Delta|\Phi(a)| \simeq \frac{\rho_m^{ini}a_{ini}^3}{a^3}(1-e^{-\Gamma_mt})(\Delta_r-\Delta_m)\,.
\end{eqnarray}
As known that $\Delta_r > \Delta_m$ during RD in the $\Lambda$CDM, $k^2\Delta|\Phi(a)|$ is expected be positive. If $\Gamma_m$ is small enough, $(1-e^{-\Gamma_mt}) \approx 0$, implying that the evolution of $|\Phi|$ is well consistent with that in the $\Lambda$CDM. Now, let us move to the short-lived scenario with a very large $\Gammadcdm$ (e.g., $\Gammadcdm=3000/30$ Mpc$^{-1}$ in Fig.~\ref{fig:Phi2}). At the RD  ($a<0.001$), Eq.~\ref{eq:Dphi_apprx} indicates the value of $|\Phi|$ in the DCDM model tends to be larger than that in the $\Lambda$CDM, and a larger $\Gammadcdm$ will lead to a larger $|\Phi|$ (i.e., $|\Phi|_{\Gammadcdm =3000}> |\Phi|_{\Gammadcdm=30}> |\Phi|_{\Lambda \rm CDM}$). However, at the epoch after the recombination ($a>0.01$), all decaying DM completely decay away for such large decay rates (i.e., $\rho_m<\rho_m^{\Lambda\rm CDM}$), giving rise to $|\Phi|< |\Phi|_{\Lambda\rm CDM}$. Along with the cosmological evolution, the trend of $|\Phi|_{\Gammadcdm =3000}> |\Phi|_{\Gammadcdm=30}$ is still observed from our numerical calculations, leading to the feature of the deeper suppression for the smaller $\Gammadcdm$ shown in Fig.~\ref{fig:Phi2}.

\acknowledgments
This work was supported by the National Science Foundation of China (11621303, 11653003, 11773021, 11835009,11890691), the National Key R\&D Program of China (2018YFA0404601, 2018YFA0404504), the 111 project, and the CAS Interdisciplinary Innovation Team (JCTD-2019-05).

\bibliographystyle{ieeetr}

\bibliography{biblio}
\end{document}